\documentclass[review,sort&compress,number]{elsarticle}

\usepackage{lineno,hyperref}
\modulolinenumbers[5]

\usepackage[utf8]{inputenc}
\usepackage{amsmath}
\usepackage{bm}
\usepackage{color}
\usepackage{siunitx}
\usepackage[bbgreekl]{mathbbol}
\usepackage{graphicx}
\usepackage{floatrow}
\usepackage[font=small,labelfont=bf]{caption}
\usepackage{multirow}
\usepackage{calligra}

\journal{Comput. Methods Appl. Mech. Engrg.}

\everymath{\displaystyle}

\definecolor{SC1}{RGB}{0,0,0}  
\definecolor{SC4}{RGB}{0,0,0}   
\definecolor{SC3}{RGB}{0,0,0}  
\definecolor{SC2}{RGB}{0,0,0}  

\newcommand{\bn}{\bm\nabla}
\newcommand{\re}{\mathrm{R}}
\newcommand{\ce}{\mathrm{C}}
\newcommand{\FMF}{\mathbf{F}^{-1}\mathbf{M}^\ast\mathbf{F}^{-\mathrm{T}}}
\newcommand{\DKDc}{\bn\cdot\mathbf{K}^\ast\bn c}

\newfloatcommand{capbtabbox}{table}[][\FBwidth]

\begin{document}

\begin{frontmatter}

\title{Phase-field study of electrochemical reactions at exterior and interior interfaces in 
Li-Ion battery electrode particles}

\author[CE,MFM]{Ying Zhao\corref{mycorrespondingauthor}}
\cortext[mycorrespondingauthor]{Corresponding author}
\ead{zhao@gsc.tu-darmstadt.de}

\author[MFM]{Bai-Xiang Xu}
\ead{xu@mfm.tu-darmstadt.de}

\author[MFM]{Peter Stein}
\ead{p.stein@mfm.tu-darmstadt.de}

\author[FM]{Dietmar Gross}
\ead{gross@mechanik.tu-darmstadt.de}

\address[CE]{Graduate School of Computational Engineering, TU~Darmstadt,\\
Dolivostrasse 15, 64293 Darmstadt,
Germany}
\address[MFM]{Mechanics of Functional Materials Division, Department of Materials Science, TU~Darmstadt,
Jovanka-Bontschits-Str. 2, 64287 Darmstadt, Germany}
\address[FM]{Solid Mechanics Division, Department of Civil Engineering, TU~Darmstadt,
Franziska-Braun-Str. 7, 64287 Darmstadt, Germany}

\begin{abstract}
To study the electrochemical reaction on 
surfaces, phase interfaces, and crack
surfaces in the lithium ion battery electrode particles,
a phase-field model is developed, which describes fracture in large strains and 
anisotropic Cahn-Hilliard-Reaction. Thereby the concentration-dependency of the elastic 
properties and the anisotropy of diffusivity are also considered. The implementation in
3D is carried out by isogeometric finite element methods
in order to treat the high order terms in a straightforward sense. 
The electrochemical reaction is modeled through a modified Butler-Volmer equation
to account for the influence of the phase change on the reaction on exterior surfaces.
The reaction on the crack surfaces is considered through a volume source term  
weighted by a term related to the fracture order parameter.
Based on the model, three characteristic examples are considered
to reveal the electrochemical reactions on particle surfaces, 
phase interfaces, and crack surfaces, as well as their influence on the particle material behavior. 
Results show that the ratio between the timescale of reaction and the diffusion can have 
a significant influence on
phase segregation behavior, as well as the anisotropy of diffusivity. In turn, the distribution of the lithium concentration
greatly influences the reaction on the surface, especially when the phase interfaces appear on exterior surfaces or crack surfaces. The reaction rate increases considerably at phase interfaces, due to the large lithium concentration gradient. 
Moreover, simulations demonstrate that the segregation of a Li-rich and a Li-poor phase during delithiation can drive the cracks to propagate.
Results indicate that the model can capture the electrochemical reaction on the freshly 
cracked surfaces. 

\end{abstract}

\begin{keyword}
Electrochemical reaction\sep Phase-field modeling of fracture \sep Cahn-Hilliard-type diffusion
\sep  Isogeometric analysis\sep Lithium-ion battery electrode particles \sep Anisotropic diffusion
\end{keyword}

\end{frontmatter}

\linenumbers

\section{Introduction}
\begin{figure}
 \centering
 \includegraphics[width = \textwidth]{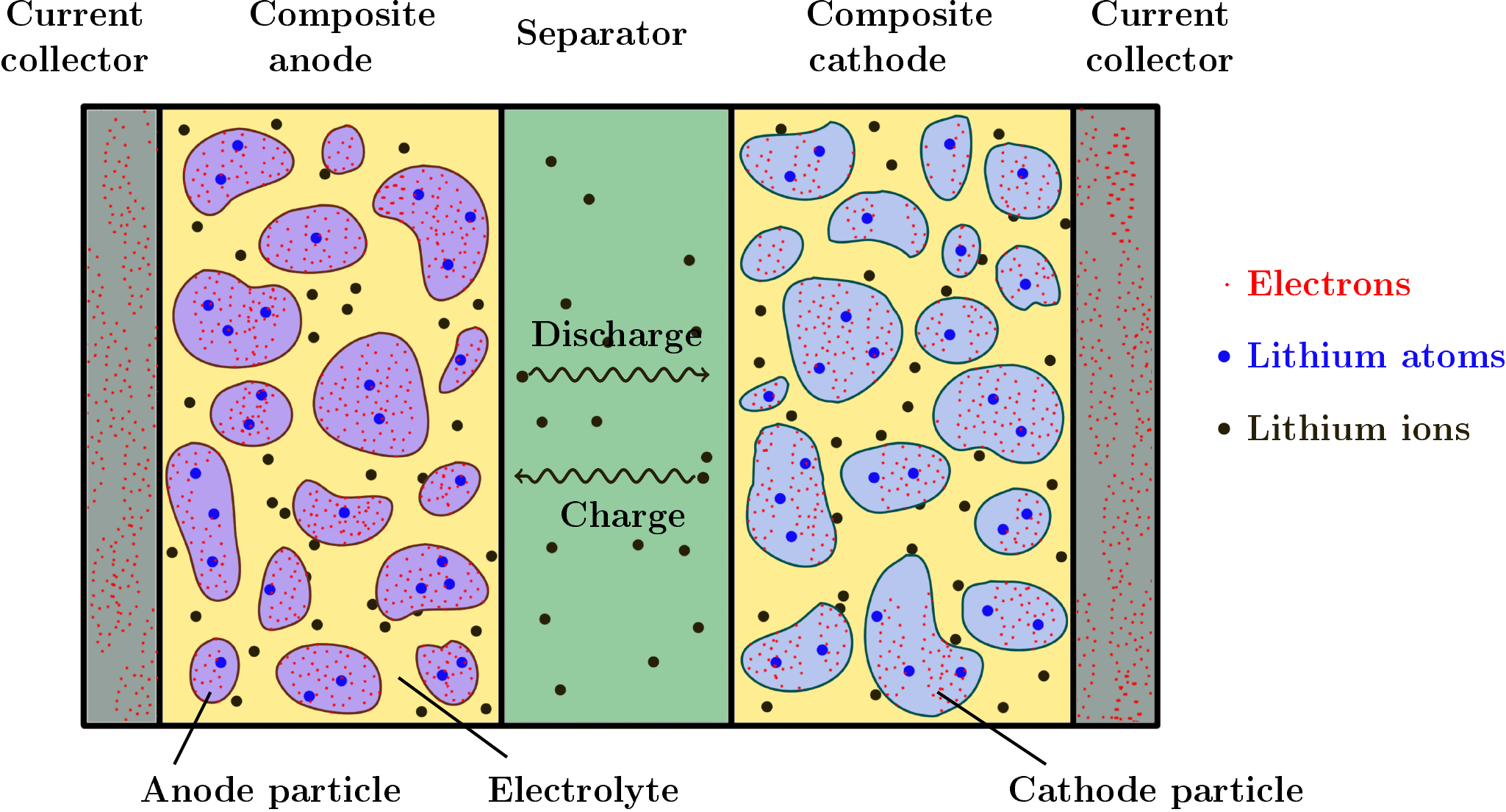}
 \caption{Schematic of a lithium-ion battery cell.}
 \label{fig:schematics}
\end{figure}

Lithium ion batteries, with their high energy densities and light-weight designs,
have found wide applications in portable electronics and electric vehicles.
A typical lithium ion battery cell is illustrated in Figure~\ref{fig:schematics}.
The current collectors and the binders between the electrodes (not depicted here) 
conduct the electrons, while the separator only permits the diffusion of lithium ions. 
The anode and cathode particles are surrounded by the electrolyte. Lithium ions intercalate
into the electrodes through electrochemical reactions on the surface of the particles.
In the electrochemical system of a battery, the reaction rate is a key point
since it is directly related to the charging/discharging performance of a 
battery. The phenomenological Butler-Volmer (BV) equation, which is based on a 
dilute solution model,
may not be able to account for a separation of phases with different Li concentrations in materials, such as silicon
and LiFePO$_4$. In the work of Singh et al.~\cite{Singh_2008}, a generalized BV
kinetics model was proposed, which includes the influence of the phase transition
on the surface reaction in a 1D case. Based on this model,
Bai et al.~\cite{Bai_2011} discussed the suppression of the phase segregation under large reaction rate.
The two dimensional case, which also coupled the Cahn-Hilliard bulk diffusion
was studied by Dargaville and Farrell~\cite{Dargaville_2013}. Using different limits of the 1D case,
they discussed when the orthotropic diffusivity becomes more isotropic.
In the mechanically coupled modeling, there has also been a tendency recently to treat the
electrochemical reaction on the surface directly through the 
BV equation rather than simply 
 to replace the reaction by a source of constant or time dependent flux~\cite{Di_Leo,Dal_computational_2014}.

The mechanical degradation of the electrode particle is widely believed to be closely 
related to the failure of the batteries and has been intensively studied in various 
chemo-mechanical coupled models~\cite{Christensen,Zhang,Stein,Huang}.  However, those models mainly treat the diffusion 
process as in a dilute solution, where the concentration smoothly changes with the incoming/outgoing flux,
accompanied by a homogeneous ``breathing-like'' expansion and shrinkage of the particle, which will hardly lead 
to the failure of the electrode particles. In the work of Huttin et al.~\cite{Huttin} and Walk et al.~\cite{Walk_2014},
Cahn-Hilliard equation was employed to investigate the stress state and compared it with the case of a dilute solution.
The diffusion process was treated as isotropic in both works.
However, as Rohrer et al.~\cite{Rohrer_insights_2013,Rohrer_origin_2015} have pointed out from first principle calculations, the anisotropic volumetric
expansion in Silicon will indeed initiate cracking, especially in large particles, where the segregation between amorphous and crystalline
silicon phases can not be suppressed. Moreover, in positive electrode materials such as LiFePO$_4$, striped phase boundaries have 
observed by Chen et al.~\cite{Chen_electron_2006,Ramana_2009} because of strong anisotropy and phase segregation. 
It demonstrates the necessity to employ a Cahn-Hilliard model and to consider the anisotropic diffusion property
coupled with large deformations in describing the bulk behavior of the particle.

The dynamics of crack propagation in lithium ion battery electrode particles has long been a challenge.
Recently, as the concept of phase-field modeling finds more applications in different disciplines, 
phase-field methods are also introduced to predict the crack
propagation coupled with diffusion. In the phase-field fracture models, the damaged and undamaged materials are considered as two different phases, indicated by the distinct values of the order parameter. Schneider et al.~\cite{Schneider_2014} proposed a model coupling the mechanics with
a general multiphase and multicomponent phase-field approach to describe the diffusion and crack propagation in brittle materials.
Liang et al.~\cite{Liang_phase_field_2014} developed a phase-field model to predict the crack evolution in LiFePO$_4$ 
cathode nanoparticles in of Li-ion batteries. Concurrently, the phase-field fracture simulation in silicon anodes is also carried
out by Zuo et al.~\cite{Zuo_2014}. Recently, Miehe et al.~\cite{Miehe_phase_2015}
conducted a comprehensive study on chemical reaction on fracture surfaces in the framework 
of phase-field fracture modeling, in addition to the reactions on exterior surfaces.

In this paper, the authors propose a phase-field model
which accounts for electrochemical reactions on different
interior and exterior of phases and crack surfaces based on the fully coupled Cahn-Hilliard-Reaction (CHR)
model proposed in the work of Bazant~\cite{Bazant_theory_2013}.
To meet the demand of higher-order continuity arising from the 
Cahn-Hilliard equation, isogeometric analysis is employed.

The paper is organized as follows. In section~\ref{sec:FractureModel}, the phase-field fracture model coupled with 
anisotropic Cahn-Hilliard-Reaction problem in large strain model is formulated.
The modeling of electrochemical reactions on the surface, phase interface, and on the crack surface is shown in 
section~\ref{sec:ElectrochimicalReaction}. Numerical details are given in section~\ref{sec:NumericalTreatments}.
Finally, three examples are given in section~\ref{sec:SimulationResults} to discuss the reaction rate, anisotropic diffusion,
and the reaction on the crack surface.

\section{Phase-field model of fracture and phase separation}\label{sec:FractureModel}
\subsection{Kinematics}\label{subsec:FM_Kinematics}
According to continuum theory, the material coordinates $\mathbf{X}$ label each
material point, while the spatial coordinates $\mathbf{x}$ denote
points in the space. The motion of the body can be described
by tracking the spatial coordinates of the material points at time $t$, i.e. $\mathbf{x} = \bm{\phi}(\mathbf{X},t)$.
The deformation gradient at a given time is then defined as 
\begin{equation}
\mathbf{F} = \bm\nabla_{\!\mathrm{R}}\,\bm\phi
\end{equation} 
in which $\bm\nabla_{\!\mathrm{R}}$ denotes
the gradient with respect to the material point $\mathbf{X}$ in the reference (material) configuration.
%
The deformation gradient is multiplicatively decomposed into two parts:
\begin{equation}
 \mathbf{F} = \mathbf{F}^e\mathbf{F}^c,
\end{equation}
where $\mathbf{F}^e$ denotes the elastic distortion,  
and $\mathbf{F}^c$ the (de-)intercalation-induced deformation.
%
The (de-)intercalation-induced deformation is usually assumed to be volumetric, and it can be further defined as
\begin{equation}
  \mathbf{F}^c = \left(J^c\right)^\frac{1}{3}\mathbf{1},
 \quad\mathrm{with}\;\; J^c = 1 + \Omega c_\mathrm{R}
 \label{eq:concentration}
\end{equation}
in which $c_\mathrm{R}(\mathbf{X},t)$ is the molar concentration
per unit volume in the reference configuration, and
$\Omega$ is the constant partial molar volume. Applying the normalization with respect to the maximum concentration $\mathrm{c}_\mathrm{max}$,
\begin{equation}
 c = {c_\mathrm{R}}/{\mathrm{c}_\mathrm{max}},\quad
 \,\Omega^\ast = {\Omega}{\mathrm{c}_\mathrm{max}}.
\end{equation}
one has
\begin{equation}
 J^c =  1 + \Omega^\ast c.
\end{equation}
On the other hand, one can also define the volumetric part of the elastic contribution $\mathbf{F}^e$, 
\begin{equation}
 J^e = \mathrm{det}\,\mathbf{F}^e = J/J^c,
 \quad\mathrm{with}\;\; J = \mathrm{det}\,\mathbf{F}.
 \label{eq:Je}
 \end{equation}
 %
The right Cauchy-Green deformation tensors are
\begin{equation}
 \mathbf{C} = \mathbf{F}^\mathrm{T}\mathbf{F},\quad
 \mathbf{C}^e = (\mathbf{F}^e)^\mathrm{T}\mathbf{F}^e=\left(J^c\right)^{-\frac{2}{3}}\mathbf{C}.
\label{eq:right_CG}
\end{equation}
%
Due to volumetric feature of $\mathbf{F}^c$, 
the deviatoric part of the total deformation
equals that of the
elastic deformation,
i.e.,
\begin{equation}
 \bar{\mathbf{C}} = \bar{\mathbf{C}}^e,
 \quad \mathrm{with}\,\,\, \bar{\mathbf{C}} = J^{-\frac{2}{3}}\mathbf{C},\quad
 \bar{\mathbf{C}}^e = (J^e)^{-\frac{2}{3}}\mathbf{C}^e.
\end{equation}
It follows that
\begin{equation}
 \bar I_1 = \bar I^e_1,
 \quad \mathrm{with}\,\,\,\bar I_1 = \mathrm{tr}(\bar{\mathbf{C}}),\quad
 \bar I^e_1 = \mathrm{tr}(\bar{\mathbf{C}}^e).
 \label{eq:I1}
\end{equation}
%
From the deformation gradient, the Green-Lagrange strain tensor is then defined as
\begin{equation}
 \mathbf{E} = \frac{1}{2}\left(\mathbf{F}^\mathrm{T}\mathbf{F} - \mathbf{1}\right)
 = \frac{1}{2}\left(\mathbf{C} - \mathbf{1}\right).
\end{equation}


\subsection{Free energy density}\label{subsec:FM_FreeEnergyDensity}

We assume a free energy coupling the chemical and the mechanical field with a damage variable $\xi$ as
\begin{equation}
 \psi_\re(c_\re,\bn_{\!\!\re} c_\re,\mathbf{C}, \xi,\bn\xi) = \psi_\re^c(c_\re) 
 + \psi_\re^i(\bn_{\!\!\re} c_\re) + \psi_\re^f(\xi,\bn_{\!\!\re}\xi)
 + \psi_\re^e(c_\re,\mathbf{C}, \xi) ,
 \label{eq:total_free_energy}
\end{equation}
where $\psi_\re^c$, $\psi_\re^i$, $\psi_\re^f$ and $\psi_\re^e$ 
are the bulk chemical free energy, the phase interface free energy, 
the fracture free energy, and the elastic free energy, respectively.
In this paper, entities with subscript R indicate those defined per volume at reference
configuration, unless otherwise indicated. 

The first two terms allow for coexistence of two phase with different Lithium concentration, and the phases are separated by a diffusive interface. The bulk free energy $\psi^c_\re$ is only dependent on the concentration $c$ and is given as
\begin{equation}
 \psi_\re^c(c) = \mathrm{RTc}_\mathrm{max}\left[c\ln c + (1-c)\ln(1-c)\right]+
             \mathrm{RTc}_\mathrm{max}\chi c(1-c),
  \label{eq:BulkFreeEnergy}
\end{equation}
in which R and T are the gas constant and the reference temperature, respectively.
To achieve a double-well function of $\psi^c$, so that this energy density allows 
for a coexistence of two phases,
one need to choose $\chi > 2$. In the simulation, $\chi = 2.5$ is adopted.

\begin{figure}[h!]
 \centering
 \includegraphics[width = \textwidth]{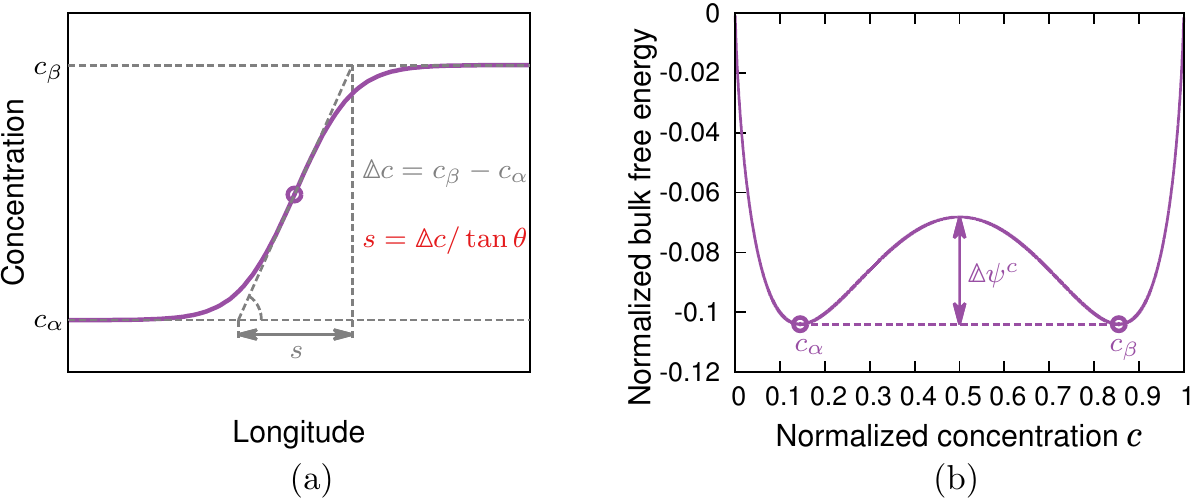}
 \caption{Definition of interface thickness for a 1D problem.}
 \label{fig:thickness}
\end{figure}

The interfacial free energy $\psi_\re^i$ gives an energetic penalty for the interface, which is expressed as
\begin{equation}
 \psi_\re^i(\bn c_\re) = \frac{1}{2}\mathrm{c}_\mathrm{max}
 {\color{SC1}\bm\nabla_\re c\cdot\mathbf{K}\bm\nabla_\re c}.
 \label{eq:InterfacialFreeEnergy}
\end{equation}
Here, $\mathbf{K}$ is an interfacial parameter, which is defined as
\begin{equation}
 \mathbf{K} = \left[
               \begin{array}{c c c}
                \kappa_x&&\\
                &\kappa_y&\\
                &&\kappa_z
               \end{array}
              \right]
\end{equation}
to exhibit an orthotropy of the interface. The parameters
$\kappa_x$, $\kappa_y$ and $\kappa_z$ are 
related to the interfacial thickness in the corresponding directions.
For a 1D case, if the elastic influence is absent, we can obtain the interfacial
thickness and the integrated interfacial energy as~\cite{Cahn_1958}
\begin{equation}
 s = \mathbb{\Delta}c/\tan\theta = (c_\beta -
 c_\alpha)\sqrt{\kappa^\ast/(2\mathbb{\Delta\psi}^c_\mathrm{max})},
 \label{eq:interface_thickness}
\end{equation}
\begin{equation}
 \Psi^i = \int_{-\infty}^\infty\frac{\psi^i_\re}{\mathrm{RTc}_\mathrm{max}} \,\mathrm{dx}
 = \int_{-\infty}^\infty \frac{1}{2}\kappa^\ast\left(c_{,x}\right)^2\,\mathrm{dx}
   = \int_{c_\alpha}^{c_\beta}\sqrt{\kappa^\ast\mathbb{\Delta}\psi^c/2}\,\mathrm{dc},
   \label{eq:interface_energy}
\end{equation}
where $s$ is the
interface thickness defined as in Figure~\ref{fig:thickness}(a),
$c_\alpha,\,c_\beta$ and
$\mathbb{\Delta\psi}^c$
are shown in Figure~\ref{fig:thickness}(b), and $\kappa^\ast = \kappa/(\mathrm{RTL}_0^2)$,
with L$_0$ being a characteristic length scale.
We can see from \eqref{eq:interface_thickness} and \eqref{eq:interface_energy} that, for a given bulk free energy, the interface
thickness and the total energy are proportional to the square root of $\kappa$, that is,
$s,\,\Psi_i\propto\sqrt{\kappa}$. It can be further concluded that,
in the 3D case, if the interfacial parameter $\kappa$ in one direction is much smaller than
those in the other two directions, the interfacial thickness and the energy expended across
the interface will be much smaller in this direction. For instance, 
$\kappa_x\ll\kappa_y=\kappa_z$ gives $s_1\ll s_2 = s_3$ and $\Psi^i_1\ll\Psi^i_2=\Psi_3$. 

The damage-like order parameter $\xi$ is introduced to describe the damage state of the material,
with a value of 1 when the material is unbroken and being 0 when it is fully broken.
According to Bourdin et al.~\cite{Bourdin_variational_2008}, the fracture free energy density is given by, 
\begin{equation}
 \psi_\re^f(\xi,\bn_{\!\!\re}\xi) = \mathcal{G}_c\left[\epsilon|\bn_{\!\!\re}\xi|^2
     + \frac{1}{4\epsilon}(1-\xi)^2\right].
  \label{eq:FractureFreeEnergy}
\end{equation}
Here, $\mathcal{G}_c$ is the critical energy release rate, and 
$\epsilon$ is a length scale which determines the width of the transition zone 
between the unbroken and the broken region.

The elastic energy $\psi^e_\mathrm{R}$ represents a stored energy of the elastic deformation. 
Although a crystal with anisotropic chemical properties will also show an anisotropy
in mechanical properties, the quantitative relation is still unknown.
For simplicity, for the undamaged region an isotropic neo-Hookean model is assumed
\begin{equation}
 \psi_\re^{e0}(c_\re,\mathbf{C}) 
   =J^c\left[\frac{K_c }{2}\left(J^e -1\right)^2
   + \frac{G_c }{2}\left(\bar{I}_1-3\right) \right],
\end{equation}
in which $K_c$ and $G_c$ are phase-dependent elastic moduli which are expressed as
\begin{equation}
 K_c = \mathrm{K}_0\left(c-\mathrm{c}_\mathrm{in}\right),
 \quad G_c = \mathrm{G}_0\left(c-\mathrm{c}_\mathrm{in}\right).
\end{equation}
Here, K$_0$, G$_0$ and c$_\mathrm{in}$ are constants, and can be determined by a linear fitting
of the measurements of the elastic moduli at different concentrations.
For more details of an undamaged isotropic model we refer the reader to our previous work~\cite{Zhao_2015}. 


Following Schneider et al.~\cite{Schneider_2014} and Zuo et al.~\cite{Zuo_2014}, 
we ignore the direct influence of diffusion
on the crack propagation. That is, the chemical field will not directly lead to fracture, but through
the stress field. Moreover, to account for the fact that cracks will not propagate under
compressive volumetric stresses, the elastic free energy can be split into a positive part $\psi_\re^{e+}$ 
and a negative part $\psi_\re^{e-}$. The latter will not be involved in the coupling with the fracture.
More specifically, the two parts take the form of
\begin{subequations}
 \begin{align}
 &\psi_\re^{e+}(c_\re,\mathbf{C}) =J^c\left[\frac{K_c }{2}\left(J^{e+} -1\right)^2
   + \frac{G_c }{2}\left(\bar{I}_1-3\right) \right], \\
 &\psi_\re^{e-}(c_\re, J) = J^c\frac{K_c}{2}\left(J^{e-}-1\right)^2,
\end{align}
\end{subequations}

in which
\begin{equation}
 \left\lbrace
 \begin{array}{c c c c}
  J^{e+} = J^e, & J^{e-}  = 1 , &\mathrm{if}& J^e \geq 1;\\
  J^{e+} =1, & J^{e-}  = J^e, &\mathrm{if}& J^e < 1.
 \end{array}
 \right.
\end{equation}
The elastic energy is defined as
\begin{equation}
 \psi_\re^e(c_\re,\mathbf{C}, \xi) = (\xi^2 + \eta)\psi_\re^{e+} + \psi_\re^{e-},
 \label{eq:ElasticFreeEnergy}
\end{equation}
in which, $0 < \eta \ll 1$ is a constant introduced to prevent singularity inside the broken
phase when $\xi = 0$. 
This method has been successfully implemented in the works of Kuhn et al.~\cite{Kuhn_2010,
Kuhn_2015} and Schl\"uter et al.~\cite{Schluter_2014} with a careful choice of $\eta$.


\subsection{Governing equations}

In this model, there are three sets of field variables: a molar concentration $c_\re$,
displacements $\bm{u}$, and a damage variable $\xi$.
The variation of the total free energy can be expressed in terms of variations of those
variables as
\begin{equation}
 \delta\Psi =  \int_{\mathrm{B}_\re}\mathbf{S}_\re:\delta\mathbf{E}\,\mathrm{dB}
  + \int_{\mathrm{B}_\re}\mu_\re\,\delta c_\re\,\mathrm{dB}
  + \int_{\mathrm{B}_\re}\zeta\,\delta\xi\,\mathrm{dB},
  \label{eq:variation1}
\end{equation}
in which $\mathbf{S}_\re$ is the second Piola-Kirchhoff stress tensor, $\mu_\re$ is the chemical potential,
$\zeta$ is the driving force for the fracture.  $\Psi$ is defined as the free energy over the whole body as
\begin{equation}
 \Psi = \int_{\mathrm{B}_\re}\psi_\re(c_\re,\bn_{\!\!\re} c_\re,\mathbf{C}, \xi,\bn\xi) \,\mathrm{dB}.
\end{equation}
With $\psi_\re$ defined in \eqref{eq:total_free_energy}, the variation of $\Psi$ is given by
\begin{align}
  \delta\Psi &= \int_{\mathrm{B}_\mathrm{R}} 
 \left[\frac{\mathrm{d}\psi^c_\mathrm{R}}{\mathrm{d}c_\mathrm{R}}\delta c_\mathrm{R}
 + { \frac{\mathrm{d}\psi^i_\mathrm{R}}
 {\mathrm{d}\bn_{\!\mathrm{R}} c_\mathrm{R}}}\cdot\delta\bn_{\!\mathrm{R}} c_\mathrm{R}
 +\frac{\partial\psi^e_\mathrm{R}}{\partial c_\mathrm{R}}\delta c_\mathrm{R}
 +\frac{\partial\psi^e_\mathrm{R}}{\partial \mathbf{C}}:\delta\mathbf{C}\right.
 \notag\\&\quad
 +\left.\frac{\partial\psi^e_\mathrm{R}}{\partial \xi}\delta\xi
 +\frac{\partial\psi^f_\mathrm{R}}{\partial \xi}\delta\xi
 +\frac{\partial\psi^f_\mathrm{R}}{\partial \bn\xi}\cdot\delta\bn\xi\right]\,\mathrm{dB}\notag\\&
 = \int_{\mathrm{B}_\mathrm{R}}\left\{\left(
 \frac{\mathrm{d}\psi^c_\mathrm{R}}{\mathrm{d}c_\mathrm{R}} 
 - \bn_{\!\re}\cdot\mathbf{K}\bn_{\!\re}c_\re
 + \frac{\partial\psi^e_\mathrm{R}}{\partial c_\mathrm{R}}\right)\delta c_\mathrm{R}
 + \bn_{\!\mathrm{R}}\cdot(\mathbf{K}\bn_{\!\mathrm{R}} c_\mathrm{R}\delta c_\mathrm{R})
 + \frac{2\partial\psi^e_\mathrm{R}}{\partial\mathbf{C}}:\delta\mathbf{E}\right.\notag \\&\quad
 +  \left[2\xi\psi^{e+}_\re
 \left.+\frac{\mathcal{G}_c}{2\epsilon}\left(\xi-1\right) 
 - 2\mathcal{G}_c\epsilon\Delta_\re\xi\right]\delta\xi
 +\bn_{\!\re}\cdot\left(\bn_{\!\re}\xi\delta\xi\right) \right\}
 \,\mathrm{dB}\notag \\
 &= \displaystyle\int_{\mathrm{B}_\mathrm{R}}\left(
 \frac{\mathrm{d}\psi^c_\mathrm{R}}{\mathrm{d}c_\mathrm{R}} -  \bn_{\!\re}\cdot\mathbf{K}\bn_{\!\re}c_\re
 + \frac{\partial\psi^e_\mathrm{R}}{\partial c_\mathrm{R}}\right)\delta c_\mathrm{R}\,\mathrm{dB}
 +\int_{\mathrm{B}_\mathrm{R}}\frac{2\partial\psi^e_\mathrm{R}}
 {\partial\mathbf{C}}:\delta\mathbf{E}\,\mathrm{dB}\notag\\&\quad
+ \int_{\mathrm{B}_\re} \left[2\xi\psi^{e+}_\re
 +\frac{\mathcal{G}_c}{2\epsilon}\left(\xi-1\right) 
 - 2\mathcal{G}_c\epsilon\Delta_\re\xi\right]\delta\xi\,\mathrm{dB}
 \notag\\&\quad
 +{\int_{\partial \mathrm{B}_\mathrm{R}}\mathbf{K}\bn_{\!\re}
 c_\mathrm{R}\cdot\mathbf{n}_\mathrm{R}\,\delta c_\mathrm{R}\,\mathrm{dS}}
 +\int_{\partial\mathrm{B}_\re}\bn_{\!\re}\xi\cdot\mathbf{n}_\re\,\delta\xi\,\mathrm{dS}.
  \label{eq:variation2}
\end{align}
Comparing \eqref{eq:variation1} and \eqref{eq:variation2},
$\mathbf{S}_\re$,  $\mu_\re$, $\zeta$ can be written as
\begin{subequations}
 \begin{align}
  &\mathbf{S}_\re = \frac{2\partial\psi_\re}{\partial\mathbf{C}}
  =\left(\xi^2+\eta\right)\frac{2\partial\psi_\re^{e+}}{\partial\mathbf{C}} 
 + \frac{2\partial\psi_\re^{e-}}{\partial\mathbf{C}},\\
 &\mu_\re = \frac{\mathrm{d}\psi_\re^c}{\mathrm{d} c_\re} + \frac{\partial\psi_\re^e}{\partial c_\re}
 - \bn_{\!\!\re}\cdot\mathbf{K}\bn_{\!\!\re}c,\label{eq:chemical_potential_define}\\
 &\zeta  = 2\xi\psi_\re^{e+}
 - 2\mathcal{G}_c\epsilon\Delta_{\!\re}\xi 
 + \frac{\mathcal{G}_c}{2\epsilon}(\xi-1)\label{eq:xi_driving_force_define},
 \end{align}
\end{subequations}
with two boundary conditions $\mathbf{K}\bn_\re c_\re\cdot\mathbf{n}_\re = 0$ and $\bn_\re \xi\cdot\mathbf{n}_\re = 0$ to be fulfilled
on the boundary surface in the reference configuration $\partial\mathrm{B}_\re$, in addition to the Dirichlet and Neumann boundary conditions
from physical constraints and fluxes.

As for the governing equation of the mechanical part, we assume a quasi-static loading, thus ignoring the inertia terms.
The governing equation for the local force balance 
in the body of the reference configuration B$_\mathrm{R}$ reads
\begin{equation}
 \bn_{\!\!\re}\cdot\mathbf{P}_\re = \mathbf{0},
 \label{eq:govern_u}
\end{equation}
where $\mathbf{P}_\re$ is the first Piola-Kirchhoff stress, defined as
\begin{equation}
 \mathbf{P} = \mathbf{F}\,\mathbf{S}_\re.
\end{equation}

Molar concentration $c_\re$ is a conserved order parameter and subject to Cahn-Hilliard-type
kinetics. In the authors' previous paper~\cite{Zhao_2015}, the species are assumed to be driven by a flux
defined in the reference configuration, which leads to simplification in finite element implementation.
In the present paper, a more physically based kinetics is employed, which defines the flux by the gradient of chemical potential
at the current configuration. More specifically,
\begin{equation}
 \frac{\partial c_\ce}{\partial t} = - \bn_\mathrm{\!\!C}\cdot\bm{j}_\ce 
 \label{eq:c_evolution_current}
\end{equation}
where the subscript C denotes quantities defined in the current configuration.

In the body B$_\mathrm{C}$,
the Cahn-Hilliard-type diffusion applies in the virgin state, while in the damaged region
no diffusion is considered. It leads to
\begin{equation}
\bm{j}_\ce = -\xi^2\bm{M}_c\bn\mu_\mathrm{R} \quad\mathrm{in}\quad\mathrm{B}_\ce,\\
\label{eq:flux1}
\end{equation}
where $\bm{M}_c$ is a mobility tensor defined as
\begin{equation}
 \bm{M}_c = c(1-c)\left[
 \begin{array}{c c c}
  \mathrm{M}_x&&\\
  &\mathrm{M}_y&\\
  &&\mathrm{M}_z
 \end{array}
 \right] = c(1-c)\mathbf{M},
\end{equation}
with $c(1-c)$ representing a degenerated mobility towards $c=0$ and $c= 1$.
On the boundary surface $\partial\mathrm{B}_\mathrm{C}$ and the surface of a
newly created crack $\Gamma_f$, the flux is dependent on the
electrochemical reaction, which follows a phenomenological Butler-Volmer equation. 
More details will follow in sections \ref{subsec:ECR_ParticleSurface} 
and \ref{subsec:ECR_CrackSurface}. Therefore, the flux is given as
\begin{equation}
\bm{j}_\ce\cdot\mathbf{n}_\ce =- \hat{j\,}_s\quad
\mathrm{on}\quad\partial\mathrm{B}_{\ce}\cup\Gamma_f.\label{eq:flux2}
\end{equation}
Based on density functional theory calculations, Rohrer et al.~\cite{Rohrer_origin_2015} have concluded that
the anisotropy of mobility is a consequence of orientation-dependent interface
energies, and that high-energy interfaces are more mobile than low-energy interfaces.
By recalling \eqref{eq:interface_energy}, it can be concluded that a higher $\kappa$ leads to
a larger $M$.
For simplicity, 
a proportional relation is assumed, that is, $\mathrm{M}_x : \mathrm{M}_y : \mathrm{M}_z 
= \kappa_x : \kappa_y : \kappa_z$.
The chemical potential $\mu_\re$ expresses the free energy change 
for adding/subtracting one mole lithium into/out of the system, thus being
the same for any configurations. 
The subscript R only indicates the fact that it is calculated 
by quantities in the reference configuration.
%
Given the condition that the total mass should be conserved in different configurations,
\begin{equation}
  \int_{\mathrm{B}_\ce}\frac{\partial c_\ce}{\partial t}\,\mathrm{d}\mathrm{B}
 =\int_{\mathrm{B}_\mathrm{R}}\frac{\partial c_\mathrm{R}}{\partial t}\,\mathrm{d}\mathrm{B},
\end{equation}
equation  \eqref{eq:c_evolution_current} can be pulled back straightforwardly
to the reference configuration as
\begin{equation}
 \frac{\partial c_\re}{\partial t} = \bn_{\!\!\re}\cdot \left[\xi^2c(1-c)
 J\mathbf{F}^{-1}\mathbf{M}\mathbf{F}^{-\mathrm{T}}\bn_{\!\!\re}\mu_\re\right]\quad\mathrm{in}
\quad \mathrm{B}_\mathrm{R}.
 \label{eq:govern_c}
\end{equation}
For later discussion, a dimensionless activity $a$ is introduced 
\begin{equation}
 \mathrm{RT}\ln a = \mu_\re,
 \label{eq:activity}
\end{equation}
and an activity coefficient $\gamma$ as a ratio $\gamma = a/c$.
Note that, when $a = c$ and thus $\gamma = 1$, this model degenerates to
an ideal dilute model.

As for a non-conserved order parameter $\xi$, the evolution equation follows
an Allen-Cahn-type equation
\begin{equation}
 \frac{\partial\xi}{\partial t} = -\mathrm{M}_\xi\zeta
 = -\mathrm{M}_\xi\left[2\xi\psi_\re^{e+}
 - 2\mathcal{G}_c\epsilon\Delta_{\!\re}\xi + \frac{\mathcal{G}_c}{2\epsilon}(\xi-1)\right]
\end{equation}
with M$_\xi$ as the mobility for the evolution of $\xi$.
Following Miehe et al.~\cite{Miehe_2010,Miehe_2015} and Borden et al.~\cite{Borden_2012}, to mimic the irreversibility of the crack, 
a strain-history field $\mathcal{H}_\mathrm{R}$ is introduced as a substitution of $\psi_\re^{e+}$,
which satisfies the Kuhn-Tucker conditions
\begin{equation}
  \mathcal{H}_\mathrm{R} \ge \psi_\re^{e+},\quad
 \dot{\mathcal{H}_\mathrm{R}}\ge 0,\quad
 \dot{\mathcal{H}_\mathrm{R}}\left(\psi_\re^{e+}-\mathcal{H}_\mathrm{R}\right) = 0.
\end{equation}
The evolution for $\xi$ then reads
\begin{equation}
 \frac{\partial\xi}{\partial t} =
 -\mathrm{M}_\xi\left[2\xi\mathcal{H}_\mathrm{R}
 - 2\mathcal{G}_c\epsilon\Delta_{\!\re}\xi + \frac{\mathcal{G}_c}{2\epsilon}(\xi-1)\right].
 \label{eq:govern_x}
\end{equation}

In summary, the governing equations for three field variables $\bm{u}$, $c$ and $\xi$
are given in \eqref{eq:govern_u}, \eqref{eq:govern_c} and \eqref{eq:govern_x}, respectively.

\section{Modeling of electrochemical reaction}\label{sec:ElectrochimicalReaction}
\subsection{Reaction on particle surfaces}\label{subsec:ECR_ParticleSurface}

\begin{figure}
\centering
 \includegraphics[width = 0.5\textwidth]{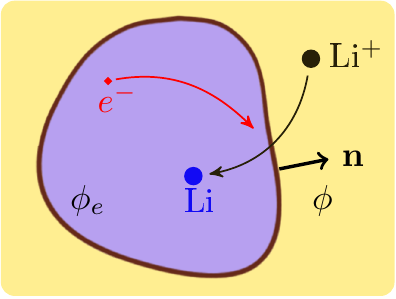}
 \caption{Illustration of the electrochemical reaction on the surface}
 \label{fig:electrochemical_reaction}
 \end{figure}

On the particle surface, a Faradaic reaction
\begin{equation}
 \mathrm{Li}^+ + e^- \rightleftharpoons \mathrm{Li}
 \label{eq:reaction}
\end{equation}
takes place, during which a Li-ion consumes an electron, as shown in Figure~\ref{fig:electrochemical_reaction}. The resultant neutral Lithium
inserts into the host material.

The rate of the reaction is described by a phenomenological Butler-Volmer (BV) 
equation,
\begin{equation}
 \hat{j\,}_s = \mathrm{c}_s\mathrm{R}_\mathrm{BV}
 = \mathrm{c}_s\frac{a_+^{1-\beta}a^\beta}{\tau_0\gamma_\mathrm{A}}
 \left[\exp\left(-\beta\frac{\mathrm{F}\eta^s_\re}{\mathrm{RT}}\right)
 -\exp\left(\left(1-\beta\right)\frac{\mathrm{F}\eta^s_\re}{\mathrm{RT}}\right)\right],
 \label{eq:butler_volmer}
\end{equation}
in which c$_s$ with
the unit of $\SI{}{\mole\per\meter\squared}$ is the molar concentration of intercalation sites on the surface, R$_\mathrm{BV}$ is the reaction rate in unit $\SI{}{\per\second}$. Moreover, $\tau_0$ is the mean time for a single reaction step, which
will be set differently to mimic a slow or fast reaction process in the simulation.
The parameter $\gamma_\mathrm{A}$ denotes the chemical activity 
coefficient of the activated state, which is taken as $(1-c)^{-1}$, while $\beta$ is a
symmetry factor for a forward and backward reaction indicated in \eqref{eq:reaction} and is set to be $0.5$. 
The Faraday constant F describes the amount of electric charge of one mole electrons.
For more details on coefficients of this model, one can refer to the work 
of Bai et al.~\cite{Bai_2011} and Dargaville et al.~\cite{Dargaville_2013}.
The definition of $a$ has been introduced in \eqref{eq:activity}. The parameters $a_+$ and $a$ are activities of Li$^+$ and Li, respectively. Since Li$^+$ diffuses
in the electrolyte much faster than Li diffuses in the electrode~\cite{Bai_2011}, $a_+$ is set to be unity
for simplicity. For a similar reason, the activity of electrons $a_-$ is also set to be 1.
The surface overpotential $\eta^s_\mathrm{R}$ is defined as the electrostatic
potential of the working electrode relative to
a reference electrode of the same kind placed in the solution adjacent to the surface of
the working electrode. It can be expressed in terms of the electrochemical potentials as
\begin{equation}\label{eq:tempt}
 \mathrm{F}\eta^s_\re = \mu_\mathrm{Li} - \mu_{\mathrm{Li}^+} - \mu_{e^-},
\end{equation}
where $\mu_\mathrm{Li}$, $\mu_{\mathrm{Li}^+}$, $\mu_{e^-}$ are electrochemical potential
of Li, Li$^+$ and $e^-$, respectively, and which are expressed as
\begin{align}
 &\mu_\mathrm{Li} = \mathrm{RT}\ln a = \mu_\re,\\
 &\mu_{\mathrm{Li}^+} = \mathrm{RT}\ln a_+ + \mathrm{F}\phi_e = \mathrm{F}\phi,\\
 &\mu_{e^-}= \mathrm{RT}\ln a_- - \mathrm{F}\phi = -\mathrm{F}\phi_e.
\end{align}
Here, $\phi_e$ denotes the electrostatic potential of the electrode, and $\phi$ represents
that of the electrolyte. Insertion of the last three equations into the surface overpotential given in \eqref{eq:tempt} leads to
\begin{equation}
 \mathrm{F}\eta^s_\re = \mu_\re + \mathrm{F}\left(\phi_e-\phi\right) 
 = \mu_\re + \mathrm{F}\mathbb{\Delta}\phi,
 \label{eq:overpotential}
\end{equation}
where $\mathbb{\Delta}\phi = \phi_e-\phi$ is the voltage drop across the electrode/electrolyte interface.
As mentioned, the subscript R in $\eta$ is only to indicate that it is expressed by quantities 
in the reference configuration and is independent from the chosen configuration.
On the other hand, the flux $\hat{j\,}_s$ is flow rate per unit area and is dependent on the
configuration. However, this dependence is fully described in the parameter c$_s$.
Therefore \eqref{eq:butler_volmer} is valid for both configurations, with the corresponding c$_s$.
The same applies for the next section, where the reaction on the newly created crack surfaces is discussed.

By substituting \eqref{eq:BulkFreeEnergy}, \eqref{eq:chemical_potential_define} 
into \eqref{eq:overpotential}, the normalized overpotential can be expressed as
\begin{equation}
 \eta^s = \frac{\mathrm{F}\eta^s_\re}{\mathrm{RT}} = \ln\frac{c}{1-c}
 + \chi(1-2c) + \mu^e - \bn\cdot\mathbf{K}^\ast\bn c+ \mathbb{\Delta}\phi^\ast,
 \label{eq:overpotential_normalized}
\end{equation}
where $\mu^e = (1/\mathrm{RT})\partial\psi^e_\re/\partial c_\re$ is the normalized
elastic chemical potential, $\mathbb{\Delta}\phi^\ast = \mathrm{F}\mathbb{\Delta}\phi/\left(\mathrm{RT}\right)$, $\mathbf{K}^\ast = \mathbf{K}/\left(\mathrm{RTL}_0^2\right)$, and $\bn = \mathrm{L}_0\bn_\re$. Here L$_0$ is a characteristic length which is introduced for normalization of the model discussed in Section~\ref{subsec:NT_Normalization}

\begin{figure}[h!]
 \centering
 \includegraphics[width = \textwidth]{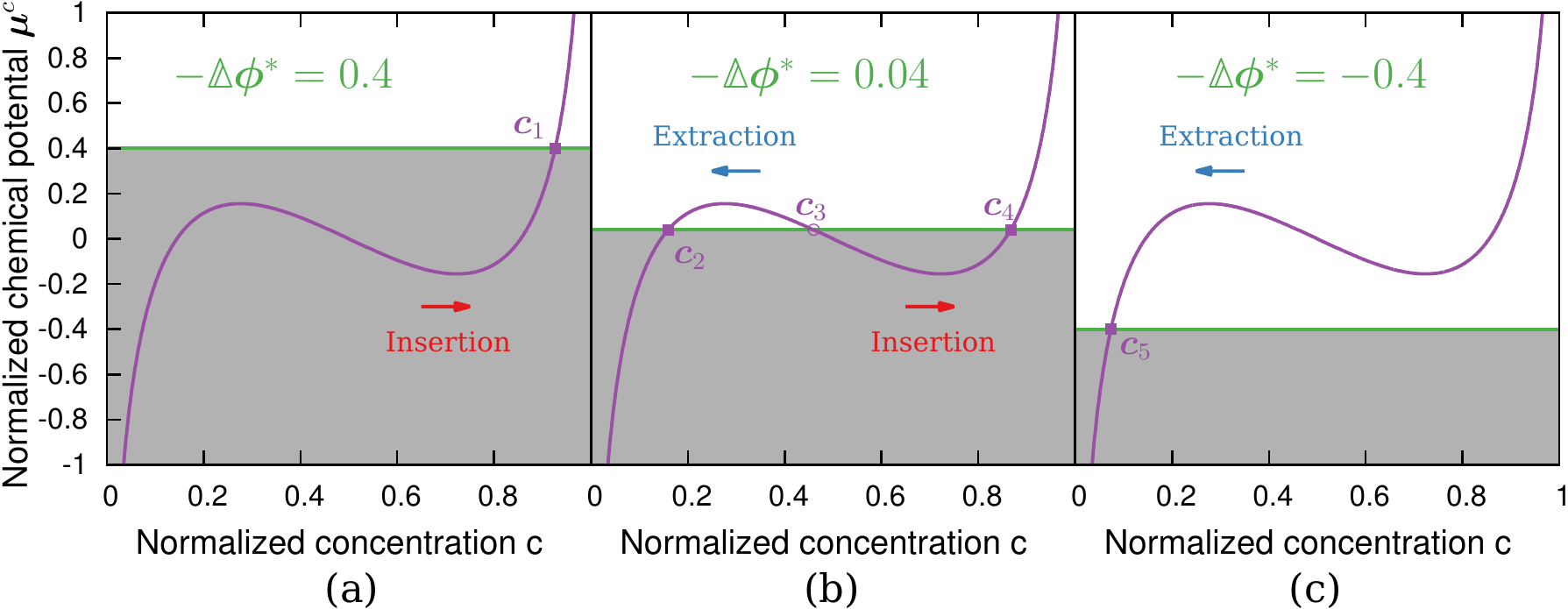}
 \caption{Insertion/extraction of Li under different normalized voltage drop 
 $\mathbb{\Delta}\phi^\ast = \mathrm{F}\mathbb{\Delta}\phi/\left(\mathrm{RT}\right)$.
 The chemical potential $\mu^c = \ln c - \ln (1-c) + 2.5\left(1-2c\right).$}
 \label{fig:Over_potential}
\end{figure}

Note that, insertion of Li takes place on the surface for $\eta^s < 0$, while extraction of Li happens for $\eta^s > 0$.
Thus, by choosing different voltage drop $\mathbb{\Delta}\phi^\ast$, the reaction  can be controlled as forward
and backward. In particular, when the interfacial and elastic chemical potential is disregarded,
$\eta^s = \mu^c + \mathbb{\Delta}\phi^\ast$. As shown in Figure~\ref{fig:Over_potential}(a), 
when $\mathbb{\Delta}\phi^\ast$ is negative and large enough,
the system will absorb Li until $c = c_1$ is reached. On the contrary,
as shown in Figure~\ref{fig:Over_potential}(c),
when $\mathbb{\Delta}\phi^\ast$ is positive and large enough, Li will be deintercalated from the system until
$c = c_5$. However, when $-\mathbb{\Delta}\phi^\ast$ stays between two spinodal points, as shown in
Figure~\ref{fig:Over_potential}(b), it is highly probable that both insertion and
extraction will take place at the same time towards $c = c_2$ and $c = c_3$, 
since the whole system is unstable due to spinodal decomposition.
Notice that $c_2$ and $c_3$ can be different from the concentrations in two phases $c_\alpha$
and $c_\beta$, which are the results of the spinodal decomposition. The values of $c_2$ and $c_3$ depend not only
on the chemical state of the material, but also on the applied voltage potential drop $\mathbb{\Delta}\phi^\ast$.
The reaction on the surface will automatically constrain the concentration in a way that the concentration
will stay in the range from 0 to 1.

Substitution of \eqref{eq:overpotential_normalized} into \eqref{eq:butler_volmer} leads to
\begin{equation}
 \hat{j\,}_s = \frac{\mathrm{c}_s}{\tau_0}\left(1-c\right)
 \exp\left(-\frac{1}{2}\mathbb{\Delta}\phi^\ast\right) 
 - \frac{\mathrm{c}_s}{\tau_0}\,c\exp\left[\chi\left(1-2c\right)+\mu^e-\bn\cdot\mathbf{K}^\ast\bn c
  +\frac{1}{2}\mathbb{\Delta}\phi^\ast\right].
  \label{eq:BV_flux}
\end{equation}


\subsection{Reaction on crack surfaces}\label{subsec:ECR_CrackSurface}

In the framework of phase-field fracture models, the crack interface can be tracked through $\xi(\xi-1) \neq 0$ or $\bn\xi\neq0$. Therefore, the boundary flux can be weighted by functions containing
either or both of these two terms, in order to consider the reaction at the crack interface.

Denote the idealized crack surface by $\Gamma_f$ (one crack will create 
two surfaces facing towards each other), and by $\Gamma^\prime_f$ the level surface in the phase-field model for a constant $\xi$.
The damage variable gradient $\bn\xi$ remains thus perpendicular
to $\Gamma^\prime_f$. Introduce a vector $\bm{s}$, which lies parallel to the damage variable gradient $\bn\xi$. 
The flux on the crack surface can hence be approximated as the flux average across the interface
\begin{equation}
 \int_{\Gamma_f} \hat{j\,}_s\,\mathrm{d}\Gamma \approx
 \frac{1}{\epsilon}\int_s\int_{\Gamma^\prime_f} g(\xi,\bn\xi)
  \hat{j\,}_s\,\mathrm{d}\Gamma^\prime\mathrm{d}s
 \approx\frac{1}{\epsilon}\int_{\mathrm{B}} g(\xi,\bn\xi)\hat{j\,}_s
 \,\mathrm{dB},
 \label{eq:approx1}
\end{equation}
in which $\epsilon$ is a length parameter and related to the interface thickness.
The weight function $g(\xi)$ contains either $\xi(\xi-1)$ or $\bn\xi$ as a factor.
If the flux is kept constant across the interface, or varies very little along the
direction of $\bn\xi$, we can observe the following relation
\begin{equation}
 \frac{1}{\epsilon}\int_s\int_{\Gamma^\prime_f} g(\xi,\bn\xi)
  \hat{j\,}_s\,\mathrm{d}\Gamma^\prime\mathrm{d}s
\approx
\frac{1}{\epsilon}\int_s g(\xi,\bn\xi)\,\mathrm{d}s
\int_{\Gamma^\prime}\hat{j\,}_s\mathrm{d}\Gamma^\prime.
 \label{eq:approx2}
\end{equation}
It follows that the approximation of \eqref{eq:approx1}
is valid, if
\begin{equation}
 \frac{1}{\epsilon}\int_s g(\xi,\bn\xi)\,\mathrm{d}s = 1.
 \label{eq:approx3}
\end{equation}

It should be commented that there will be several factors that can influence the accuracy of 
the approximation.
\begin{itemize}
 \item The phase-field approximation of the cracked phase interface will determine
 the choice of $g(\xi)$. Usually $g(\xi)$ is chosen  based on an uncoupled model for simplicity.
 However, when the mechanical stresses come into play, the $\xi$ profile
 can be different. The error can be even larger  when  $\xi(\xi-1)$ is used as its weighting term, 
 since $\xi$ might not be 0 and 1
 in two homogeneous phases, depending on the boundary condition given (the reader is referred to \cite{Borden_2012} for more details).
 Therefore $g(\xi)$ needs to be modified accordingly, when the influence of
 stresses is not negligible;
 \item The flux can vary strongly across the phase interface, especially when the diffusion is
 so slow that the concentration varies greatly along the direction of $\bn\xi$, making electrochemical
 reaction on/in the interface highly fluctuating in a small range. In these cases the equation \eqref{eq:approx2} may not be accurate enough. One can, for instance, increase the
 polynomial order of $g(\xi)$, so that the fluctuation of the reaction 
 will become negligible compared to $g(\xi)$. 
\end{itemize}

As a simple case, we set $g(\xi) = \mathrm{A}\xi^2(1-\xi)^2$ with A being a coefficient
to be determined.
To this end, firstly, the profile of $\xi(x)$ across the interface on one side can be obtained
by solving the following uncoupled 1D problem at equilibrium
\begin{equation}
\left\lbrace
\begin{array}{l c r}
 0 = 2\mathcal{G}_c\epsilon\xi^{\prime\prime}
 + \frac{1}{2}\frac{\mathcal{G}_c}{\epsilon}\left(s-1\right)
 &\mathrm{for}& 0 < x \le\mathrm{L},\\
 \xi|_{x = 0} = 0, && \\
\xi^\prime|_{x = \mathrm{L}} = 0.  &&
\end{array}
\right.
\end{equation}
The solution reads
\begin{equation}
 \xi(x) = 1 - \cosh\left(\frac{x}{2\epsilon}\right) 
 + \coth\left(\frac{\mathrm{L}}{2\epsilon}\right)\sinh\left(\frac{x}{2\epsilon}\right)
 = 1 + \sinh^{-1}\left(\frac{\mathrm{L}}{2\epsilon}\right)\sinh\left(\frac{x-\mathrm{L}}{2\epsilon}\right).
 \label{eq:ksi}
\end{equation}
Inserting \eqref{eq:ksi} into $g(\xi)$ and integrating over the whole length L, one has
\begin{align}
1&=\frac{1}{\epsilon}\mathrm{A}\int_0^\mathrm{L} \xi^2\left( 1-\xi\right)^2\,\mathrm{d}x\notag\\
&=\frac{1}{\epsilon}\mathrm{A}\int_0^\mathrm{L} \left[ 1 
 + \sinh^{-1}\left(\frac{\mathrm{L}}{2\epsilon}\right)\sinh\left(\frac{x-\mathrm{L}}{2\epsilon}\right)\right]^2
 \left[-\sinh^{-1}\left(\frac{\mathrm{L}}{2\epsilon}\right)\sinh\left(\frac{x-\mathrm{L}}{2\epsilon}\right)\right]^2
 \,\mathrm{d}x\notag\\
 &\approx \frac{1}{\epsilon}\mathrm{A}\cdot\frac{1}{6}\epsilon =\frac{\mathrm{A}}{6},
\end{align}
which gives A$=6$ and $g(\xi) = 6\xi^2(1-\xi)^2$.
The evolution for the concentration can then be expanded from 
\eqref{eq:govern_c} as
\begin{equation}
 \frac{\partial c_\re}{\partial t} = \bn_{\!\!\re}\cdot \left[\xi^2c(1-c)
 J\mathbf{F}^{-1}\mathbf{M}\mathbf{F}^{-\mathrm{T}}\bn_{\!\!\re}\mu_\re\right]
 +\frac{6}{\epsilon}\xi^2\left(1-\xi\right)^2\hat{j\,}_s,
 \label{eq:govern_c_extended}
\end{equation}
in which $\hat{j\,}_s$ is given in \eqref{eq:BV_flux} to account for the flux due to electrochemical reaction.

\section{Numerical treatment}\label{sec:NumericalTreatments}
\subsection{Normalization}\label{subsec:NT_Normalization}

For the convenience of finite element implementation, the model presented above is normalized first. Introduce a dimensionless form of space and time as
\begin{equation}
  \bar{\mathbf{X}} = \frac{\mathbf{X}}{\mathrm{L}_0},\quad
  \bar t = \frac{\mathrm{D}}{\mathrm{L}_0^2}\,t.
 \end{equation}
in which L$_0$ is a characteristic length scale, which is identical in the three directions,
and D is a diffusion coefficient of one direction.
Energy density is scaled by RTc$_\mathrm{max}$, and the other quantities 
can be normalized accordingly as
\begin{equation}
 \psi = \frac{\psi_\mathrm{R}}{\mathrm{RT}\mathrm{c}_\mathrm{max}},\quad
 \mu = \frac{\mu_\mathrm{R}}{\mathrm{RT}},\quad
 \mathbf{j} = \frac{\mathrm{L}_0}{\mathrm{D}\mathrm{c}_\mathrm{max}}\,\mathbf{j}_\mathrm{R},\quad
 \mathbf{S} = \frac{\mathbf{S}_\mathrm{R}}{\mathrm{RT}\mathrm{c}_\mathrm{max}},\quad
 \mathcal{H} = \frac{\mathcal{H}_\mathrm{R}}{\mathrm{RTc}_\mathrm{max}}.
\end{equation}
For the fracture model, the normalized fracture length scale, energy release rate and
the mobility are
\begin{equation}
 \epsilon^\ast = \frac{\epsilon}{\mathrm{L}_0},\quad
 \mathcal{G}_c^\ast = \frac{\mathcal{G}_c}{\mathrm{L}_0\mathrm{RTc}_\mathrm{max}},\quad
 \mathrm{M}_\xi^\ast = \frac{\mathrm{M}_\xi\mathrm{L}_0^2\mathrm{RTc}_\mathrm{max}}{\mathrm{D}}
\end{equation}
For the reaction, the surface site concentration c$_s$ and the single reaction time step $\tau_0$ is
\begin{equation}
 \mathrm{c}_s^\ast = \frac{\mathrm{c}_s}{\mathrm{c}_\mathrm{max}\mathrm{L}_0},\quad
\tau_0^\ast = \frac{\mathrm{D}}{\mathrm{L}_0^2}\tau_0
\end{equation}
Thus the normalized governing equations can be summarized as:
\begin{equation}
 \left\lbrace
 \begin{array}{l c l l l}
  \bn\cdot\mathbf{P} = \mathbf{0} & \mathrm{in}
  & \mathrm{B} & \times & \left[0,\bar{\mathcal{T}}\right],\\
  \dot c = \bn\cdot\left[\xi^2c(1-c)J\mathbf{F}^{-1}
  \mathbf{M}^\ast\mathbf{F}^{-\mathrm{T}}\bn\mu\right] + \frac{6}{\epsilon^\ast}
  \xi^2\left(1-\xi\right)^2\bar{\hat{j\,}}_s,
  &\mathrm{in}& \mathrm{B} & \times & \left[0,\bar{\mathcal{T}}\right],\\
  \dot\xi = - M_\xi^\ast\left[2\xi\mathcal{H}-2\mathcal{G}_c^\ast\epsilon^\ast\Delta\xi
  + \frac{\mathcal{G}_c^\ast}{2\epsilon^\ast}\left(\xi-1\right)\right],
  &\mathrm{in}& \mathrm{B} & \times & \left[0,\bar{\mathcal{T}}\right],\\

  \bar{\mathbf{u}} = \hat{\bar{\mathbf{u}}}&\mathrm{on}
  &\mathrm{S}_{\bar{\mathbf{u}}} &  \times & \left[0,\bar{\mathcal{T}}\right],\\
  \mathbf{P}\cdot\mathbf{n} = \hat{\bar{\mathbf{t}}}&\mathrm{on}
  &\mathrm{S}_{\bar{\mathbf{t}}}&  \times & \left[0,\bar{\mathcal{T}}\right],\\
  \mathbf{j}\cdot\mathbf{n} = -\bar{\hat{j\,}}_s &\mathrm{on}
  &\partial\mathrm{B}&  \times & \left[0,\bar{\mathcal{T}}\right],\\
  \mathbf{K}\bn c\cdot\mathbf{n} = 0&\mathrm{on}
  &\partial\mathrm{B}&  \times & \left[0,\bar{\mathcal{T}}\right],\\
  \bn \xi\cdot\mathbf{n} = 0&\mathrm{on}
  &\partial\mathrm{B}&  \times & \left[0,\bar{\mathcal{T}}\right],\\
  c\left(\bar{\mathbf{X}},0\right) = c_0\left(\bar{\mathbf{X}}\right)&\mathrm{in}
  &\mathrm{B},&&\\
  \xi\left(\bar{\mathbf{X}},0\right) = 0&\mathrm{on}
  &\Gamma_f,&&\\
  \xi\left(\bar{\mathbf{X}},0\right) = 1&\mathrm{in}
  &\mathrm{B}\backslash\Gamma_f,&&
 \end{array}
 \right.
 \label{eq:NorGovern}
\end{equation}
with $\bar{\hat{j\,}}_s$ defined as
\begin{equation}
 \bar{\hat{j\,}}_s = \frac{\mathrm{c}_s^\ast}{\tau_0^\ast}\left(1-c\right)
 \exp\left(-\frac{1}{2}\mathbb{\Delta}\phi^\ast\right) 
 - \frac{\mathrm{c}_s^\ast}{\tau_0^\ast}\,c\exp\left[\chi\left(1-2c\right)+\mu^e-\bn\cdot\mathbf{K}^\ast\bn c
  +\frac{1}{2}\mathbb{\Delta}\phi^\ast\right]
\end{equation}

\subsection{Implementation details}\label{subsec:NT_ImplementationDetails}

This model is implemented by using the finite element program FEAP~\cite{Taylor_feap_2014} with Non-Uniform Rational B-Splines (NURBS) as shape functions for the spatial discretization,
which allow for a straightforward treatment of the fourth-order Cahn-Hilliard 
equation. The displacements $\bm u$, the concentration $c$ and the order parameter 
$\xi$ are nodal degrees of freedom. In addition, to deal with the additional boundary condition 
$\mathbf{K}\bn c\cdot\mathbf{n} = 0$ arising along with the Cahn-Hilliard equation, a Lagrange multiplier
$\lambda$ is introduced as an additional degree of freedom for each node.
For more details about $\lambda$ the reader is referred to our previous work~\cite{Zhao_2015}.
A backward Euler method is employed for the time integration, and Newton-Raphson iteration scheme
is used for the nonlinear system of equations at each time step.

The above mentioned 6 field variables are interpolated under an isoparametric/isogeometric concept as
\begin{equation}
 \bm u = N^\mathrm{I}\bm{u}^\mathrm{I}, \quad c = N^\mathrm{I}c^\mathrm{I}, \quad 
 \xi = N^\mathrm{I}\xi^\mathrm{I}, \quad \lambda = N^\mathrm{I}\lambda^\mathrm{I},
\end{equation}
where $\left(\cdot\right)^\mathrm{I}$ is the value at the I-th control point, and $N^\mathrm{I}$
is the NURBS shape function associated with the I-th control point. The repeated I invokes the Einstein
summation. The gradient terms are thus given by
\begin{equation}
 \delta\mathbf{\underline{E}} = \mathbf{B}_{\bm u}^\mathrm{I}\delta\bm u^\mathrm{I},\quad
 \bn c = \mathbf{B}_c^\mathrm{I},\quad
 \bn \xi = \mathbf{B}_\xi^\mathrm{I},\quad
 \bn \lambda = \mathbf{B}_\lambda^\mathrm{I},
\end{equation}
where $\mathbf{\underline{E}}$ is the Green-Lagrangian strain tensor in Voigt notation,
\begin{equation}
 \mathbf{B}_c^\mathrm{I} = \mathbf{B}_\xi^\mathrm{I} = \mathbf{B}_\lambda^\mathrm{I}
 =\bn N^\mathrm{I}=
 \left[\begin{array}{c c c}
        N^\mathrm{I}_{,1}&N^\mathrm{I}_{,2}&N^\mathrm{I}_{,3}
       \end{array}
\right]^\mathrm{T},
\end{equation}
and
\begin{equation}
 \mathbf{B}_{\bm u}^\mathrm{I} = \left[
 \begin{array}{c c c}
     F_{11}N^\mathrm{I}_{,1} & F_{21}N^\mathrm{I}_{,1} & F_{31}N^\mathrm{I}_{,1}\\
    F_{12}N^\mathrm{I}_{,2} & F_{22}N^\mathrm{I}_{,2} & F_{32}N^\mathrm{I}_{,2}\\
  F_{13}N^\mathrm{I}_{,3} & F_{23}N^\mathrm{I}_{,3} & F_{33}N^\mathrm{I}_{,3}\\
     F_{11}N^\mathrm{I}_{,2}+F_{12}N^\mathrm{I}_{,1} & F_{21}N^\mathrm{I}_{,2}+F_{12}N^\mathrm{I}_{,1}
         & F_{31}N^\mathrm{I}_{,2}+F_{12}N^\mathrm{I}_{,1}\\
     F_{12}N^\mathrm{I}_{,3}+F_{13}N^\mathrm{I}_{,2} & F_{22}N^\mathrm{I}_{,3}+F_{13}N^\mathrm{I}_{,2}
       & F_{32}N^\mathrm{I}_{,3}+F_{13}N^\mathrm{I}_{,2}\\
    F_{13}N^\mathrm{I}_{,1}+F_{11}N^\mathrm{I}_{,3} & F_{23}N^\mathrm{I}_{,1}+F_{11}N^\mathrm{I}_{,3}
     & F_{33}N^\mathrm{I}_{,1}+F_{11}N^\mathrm{I}_{,3}\\                                     
 \end{array}
\right].
\end{equation}
Here, $N^\mathrm{I}_{,i}$ denotes $\partial N^\mathrm{I}/\partial X_i$, and $F_{ij}$ are
the components of the  deformation gradient $\mathbf{F}$.

Thus the discretized weak statement of \eqref{eq:NorGovern} reads
\begin{equation}
 \delta\Pi =  (\delta\mathbf{u}^\mathrm{I})^\mathrm{T}\,\mathbf{R}^\mathrm{I}_u
 + \delta c^\mathrm{I}\,\mathrm{R}^\mathrm{I}_c
 + \delta\xi^\mathrm{I}\,\mathrm{R}^\mathrm{I}_\xi
 + \delta\lambda^\mathrm{I}\,\mathrm{R}^\mathrm{I}_\lambda = 0,
\end{equation}
in which the residuals are
\begin{subequations}
 \begin{align}
   \mathbf{R}^\mathrm{I}_u
     &=\displaystyle  -\int_\mathrm{B}(\mathbf{B}^\mathrm{I}_u)^\mathrm{T}
    \left[\left(\xi^2 + \eta\right)\frac{\partial\psi^{e+}}{\partial\mathbf{\underline{E}}}
    +\frac{\partial\psi^{e-}}{\partial\mathbf{\underline{E}}}\right]\,\mathrm{dB},\\
   \mathrm{R}^\mathrm{I}_c
     &=  \int_\mathrm{B}\dot c\,N^\mathrm{I}\,\mathrm{dB}
           + \int_\mathrm{B}\xi^2 \left[1 - 2\chi c\left(1-c\right)\right]J\,
           (\bn c\cdot \FMF\bn N^\mathrm{I})\,\mathrm{dB}\notag\\
     &+ \int_\mathrm{B} \xi^2c\left(1-c\right)J\,(\bn\mu^e\cdot \FMF\bn N^\mathrm{I})
        \,\mathrm{dB}\notag\\
      &   + \int_\mathrm{B} 2\xi c\left(1-c\right)J(\DKDc)\,(\bn\xi\cdot
         \FMF\bn N^\mathrm{I})\,\mathrm{dB}\notag\\
        & + \int_\mathrm{B}\xi^2\left(1-2c\right)J\,(\DKDc)\,(\bn c\cdot
         \FMF\bn N^\mathrm{I})\,\mathrm{dB}\notag\\
         &+\int_\mathrm{B}\xi^2c\left(1-c\right)(\DKDc)\,(\bn J\cdot
         \FMF\bn N^{-1})\,\mathrm{dB}\notag\\
        & + \int_\mathrm{B}\xi^2 c\left(1-c\right)\,(\DKDc)\,
           \Delta N^\mathrm{I}\,\mathrm{dB}\notag\\
        &+\int_\mathrm{B}\xi^2c\left(1-c\right)J\,(\DKDc) \,
        \left[\left(\bn\cdot\FMF\right)\cdot\bn N^\mathrm{I}\right]\,\mathrm{dB}\notag\\
        &+\int_\mathrm{B}\xi^2c\left(1-c\right)J\,(\DKDc)
        \,\left(\FMF:\bn\bn N^\mathrm{I}\right)\,\mathrm{dB}\notag\\
        &-\int_\mathrm{B}\frac{6\xi^2\left(1-\xi\right)^2}{\epsilon^\ast}\hat{j\,}_s
        N^\mathrm{I}\,\mathrm{dB} - 
        \int_{\partial\mathrm{B}}\hat{j\,}^s\,N^\mathrm{I}\,\mathrm{dS}\notag\\
         &+\int_\mathrm{B}\bn\lambda\cdot\mathbf{K}^\ast\bn N^\mathrm{I}\,\mathrm{dB}
         + \int_\mathrm{B}\lambda\, \bn\cdot\mathbf{K}^\ast\bn N^\mathrm{I}\,\mathrm{dB},\\
   \mathrm{R}^\mathrm{I}_\xi
&=\int_\mathrm{B}\dot\xi\, N^\mathrm{I}\,\mathrm{dB}
    - \int_\mathrm{B}\mathrm{M}_\xi^\ast\left[2\xi\,\mathcal{H}+\frac{\mathcal{G}_c^\ast}{2\epsilon^\ast}
    \left(\xi-1\right)\right]N^\mathrm{I}\,\mathrm{dB}\notag\\
    &- \int_\mathrm{B}\mathrm{M}_\xi^\ast\cdot2\mathcal{G}_c^\ast\epsilon^\ast\bn\xi\cdot\bn N^\mathrm{I}
    \,\mathrm{dB}\\
    \mathrm{R}^\mathrm{I}_\lambda
     &=\int_\mathrm{B} \DKDc\,N^\mathrm{I}\,\mathrm{dB}
                 +\int_\mathrm{B} \mathbf{K}^\ast\bn c\cdot\bn N^\mathrm{I}\,\mathrm{dB}
                 -\frac{1}{\alpha}\int_\mathrm{B}\lambda\, N^\mathrm{I}\,\mathrm{dB}.
 \end{align}
Construction of the corresponding tangent matrices can be achieved according to the finite element theory. 
 
\end{subequations}

\section{Simulation results}\label{sec:SimulationResults}
\subsection{Reaction on the particle surface}\label{subsec:SR_ParticleSurface}
To study the phase segregation under different diffusion and reaction limits,
a sphere with isotropic material is considered, where a homogeneous initial and
boundary setup is also given. Symmetric mechanical constraints are applied on the
planes of symmetry, while the spherical surface is set free from stresses.
Electrochemical reaction takes place on the surface of the sphere, 
across which a constant voltage drop 
$\mathbb{\Delta\phi}$ is prescribed. It drives the reaction, such that the neutral 
Lithium is produced (consumed), until the particle is fully (dis-)charged.
The reaction rate is controlled by the single reaction step time $\tau_0$, which is
given as $\SI{0.01}{\second}$ for a fast reaction and $\SI{1}{\second}$ for a slow reaction.
The parameters for the simulation are given in Table~\ref{tb:parameters}.
\begin{table}[h!]
 \centering
 \scriptsize
 \begin{tabular}{r l}
 \hline
 \hline
  {Gas constant} (R)&$\SI{8.32}{\joule\per\mole\per\kelvin}$\\ \hline
 { Absolute temperature} (T)&$\SI{283}{\kelvin}$\\ \hline
 {Diffusivity} (D)&$\SI{7.08d-15}{\meter\squared\per\second}$\\ \hline
 { Faraday's constant} (F)&$\SI{96485}{\coulomb\per\mole}$\\\hline
 {Partial molar volume} ($\Omega$)&$\SI{3.497d-6}{\meter\cubed\per\mole}$\\ \hline
 Maximum concentration ($\mathrm{c}_\mathrm{max}$)&$\SI{2.29d4}{\mole\per\meter\cubed}$\\\hline
 Phase parameter ($\chi$)&$2.5$\\ \hline
 Interface parameter ($\kappa$)&$\SI{1.0d-10}{\joule\per\mole\meter\squared}$\\\hline
 Length scale (L$_0$)&$\SI{1}{\micro\meter}$\\\hline
 Bulk modulus slope (K$_0$)&$\SI{100}{\mega\pascal}$\\ \hline
 Shear modulus slope (G$_0$)&$\SI{100}{\mega\pascal}$\\ \hline
 Concentration intercept ($\mathrm{c}_\mathrm{in}$)&$10.0$\\\hline
 Surface site concentration (c$_s$)&$\SI{6.78d-6}{\mole\per\meter\squared}$\\\hline
 Single reaction step time ($\tau_0$)&$\SI{0.01}{\second}$ (fast)/$\SI{1}{\second}$ (slow)\\\hline
 Voltage drop electrode/electrolyte ($\mathbb{\Delta}\phi$)&$\SI{-4.88}{\milli\volt}$\\\hline
 Initial normalized concentration ($c_0$)&0.25\\\hline\hline
 \end{tabular}
 \caption{Parameters for the simulation of the spherical particle.}
 \label{tb:parameters}
\end{table}

\begin{figure}[h!]
 \centering
 \includegraphics[width = \textwidth]{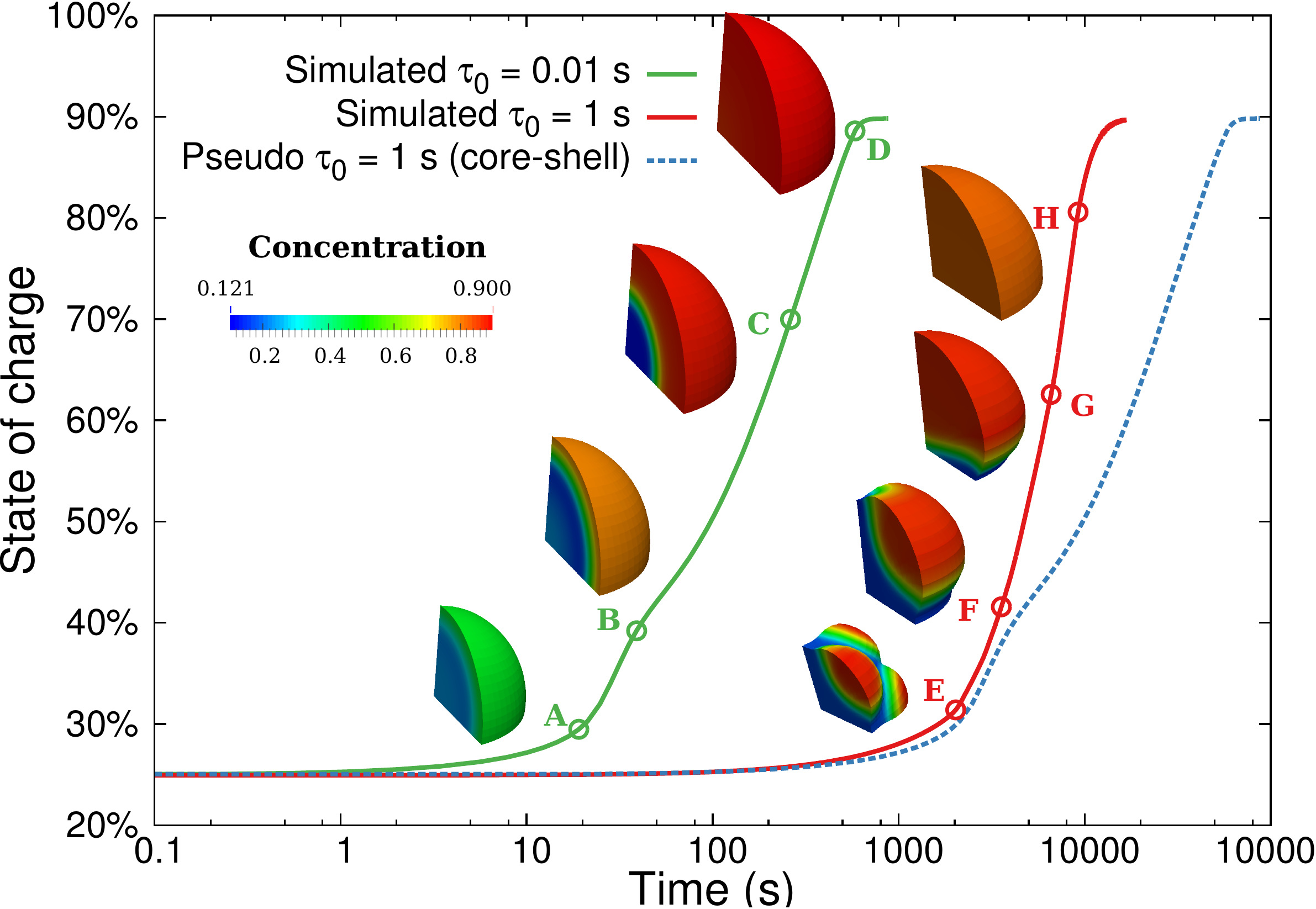}
 \caption{The SOC with corresponding concentration profile of the particle
 at different charging rate. A fast reaction (green line) can give a core-shell structure while a
 slow reaction (red line) will not preserve the core-shell. However, the latter can give a more
 robust reaction after the phase segregation occurs. The pseudo plot is the curve when
 the core-shell structure is enforced when $\tau_0 = \SI{1}{\second}$.}
 \label{fig:state_of_charge}
\end{figure}

The state of charge (SOC) with respect to time is measured in the simulation by integrating all the Lithium inside 
the particle at each current time compared with the amount in a full lithiation ($c_\mathrm{R} = \mathrm{c}_\mathrm{max}$). 
The results are shown in Figure~\ref{fig:state_of_charge}.
The solid lines describe the simulated SOC with respect to time. 
Both curves show the same tendency of the reaction,
which is fairly fast at the beginning and slows down towards the end of the charge.
Both of them show an acceleration of reaction when the particle 
is charged at roughly $\SI{30}{\percent}$ (A, E), because in both cases the phases start to 
form and the overpotential $\eta^s$ increases rapidly as the concentration increases, which can be
seen in Figure~\ref{fig:Over_potential} (a).

However, the phase segregation is very different in the two cases.
The green line shows that, when the reaction is fast enough,
a core-shell structure can be achieved. This is in agreement with the predictions in
the work of Singh et al.~\cite{Singh}, which stated that 
in an isotropic bulk-transport-limited case, where the bulk diffusion is much 
slower than the reaction, the phase boundary is driven largely by the incoming flux,
thus a shrinking-core profile being formed. On the other hand, as shown by the red line,
when the surface reaction is slow enough, the species can be always equilibrated
by the bulk transport. In this case, the spinodal decomposition, or nucleation,
initiates from the surface, where the dynamics of reaction can greatly fluctuate
the species concentration. They are very unstable inside the spinodal region.
The reaction in the core-shell structure slows down as the two phase region is finally formed (B, C),
because the outer shell approaches a full lithiation. However, in the other case, the 
reaction maintains its rate (F, G) until the phase segregation is suppressed.
Based on the simulation results of the case $\tau_0 = \SI{0.01}{s}$, 
one can predict the state of charge curve for $\tau_0 = 1 s$ when the core-shell structure in enforced, simply by scaling the time
of the fast case by a factor of 100. For comparison, this predicted result is shown by the curve in blue color in Fig.~\ref{fig:state_of_charge}. It shows that if the core-shell structure is maintained, the lithiation process becomes slower than that in the case
when the particle is free to adjust the phase pattern for a more robust reaction.

\subsection{Reaction inside the interface}\label{subsec:SR_Interphase}

\begin{figure}
\begin{floatrow}
\ffigbox{%
  \includegraphics[width =0.5\textwidth]{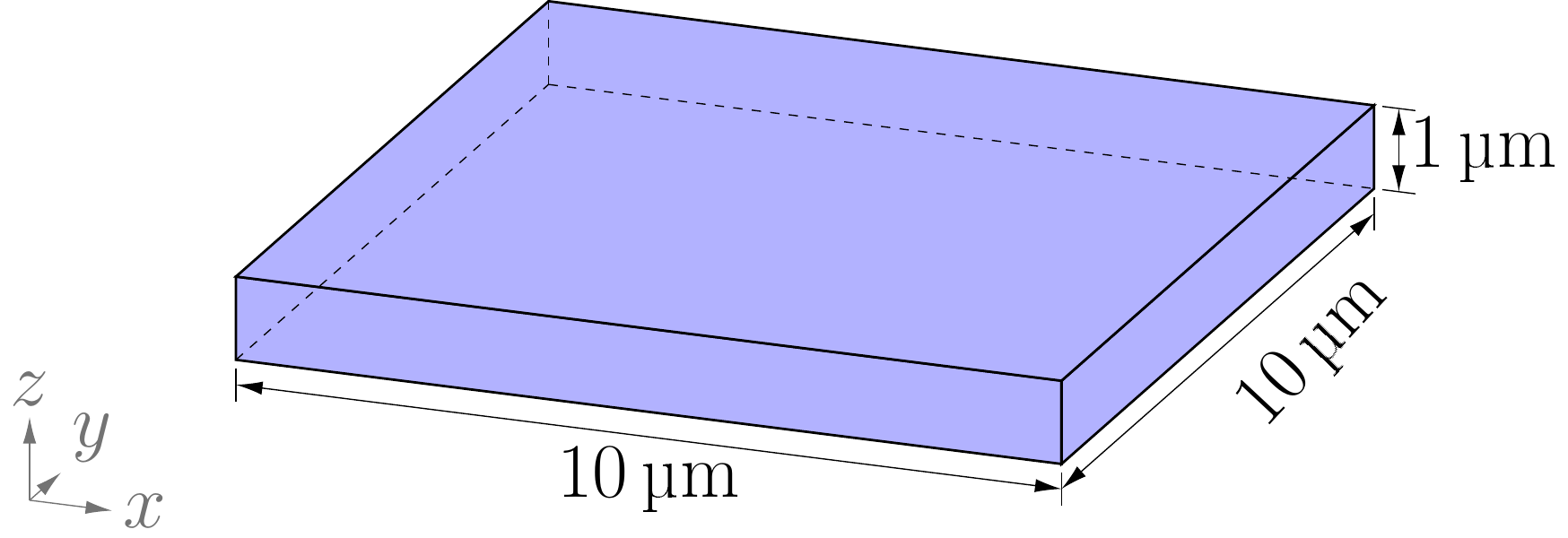}%
}{%
  \caption{Sketch and measurements of the square plate problem. All six surfaces of the plate
 are exposed to the electrolyte, therefore electrochemical reaction can take place on all sides.}%
\label{fig:model_squareplate}}
\capbtabbox{%
 \scriptsize
    \begin{tabular}{ccl}\hline\hline
      \multirow{3}{*}{\begin{tabular}{c}
                       Interfacial\\
                       Parameter
                      \end{tabular}} & $\kappa_x$ & $\SI{1.0d-10}{\joule\per\mole\per\meter\squared}$ \\ 
                                     & $\kappa_y$ & $1000\,\kappa_x$\\
                          & $\kappa_z$ & $1000\,\kappa_x$\\\hline
                  & D$_x$ & $\SI{7.08d-15}{\meter\squared\per\second}$ \\
      Diffusivity & D$_y$ & 1000 D$_x$\\
                  & D$_z$ & 1000 D$_x$\\\hline
     Sing. reac. & $\tau_0$&$\SI{1}{\second}$\\\hline\hline
      \end{tabular}
}{%
  \caption{Anisotropic interface and diffusivity parameters for the plate problem. For the
      isotropic case, the parameters for x direction are used for all three directions. All other parameters are given in Table~\ref{tb:parameters}.}%
\label{tb:parameter_SP}}
\end{floatrow}
\end{figure}

To investigate the reaction in different phases and the phase interface, a square plate
with isotropic and anisotropic chemical properties is studied in this section.
The geometry is given in Figure~\ref{fig:model_squareplate}. No preferred direction
for the reaction is assumed. Therefore, in the simulation, the reaction will take place in all six surfaces 
of the plate and the reaction rate is governed by the chemical state at each position on the
surface. The anisotropic interfacial parameter $\kappa$ and diffusivity D
in different directions is given in Tabel~\ref{tb:parameter_SP}.
In this model, the mechanical  part is disregarded.

\begin{figure}[h!]
 \centering
 \includegraphics[width = \textwidth]{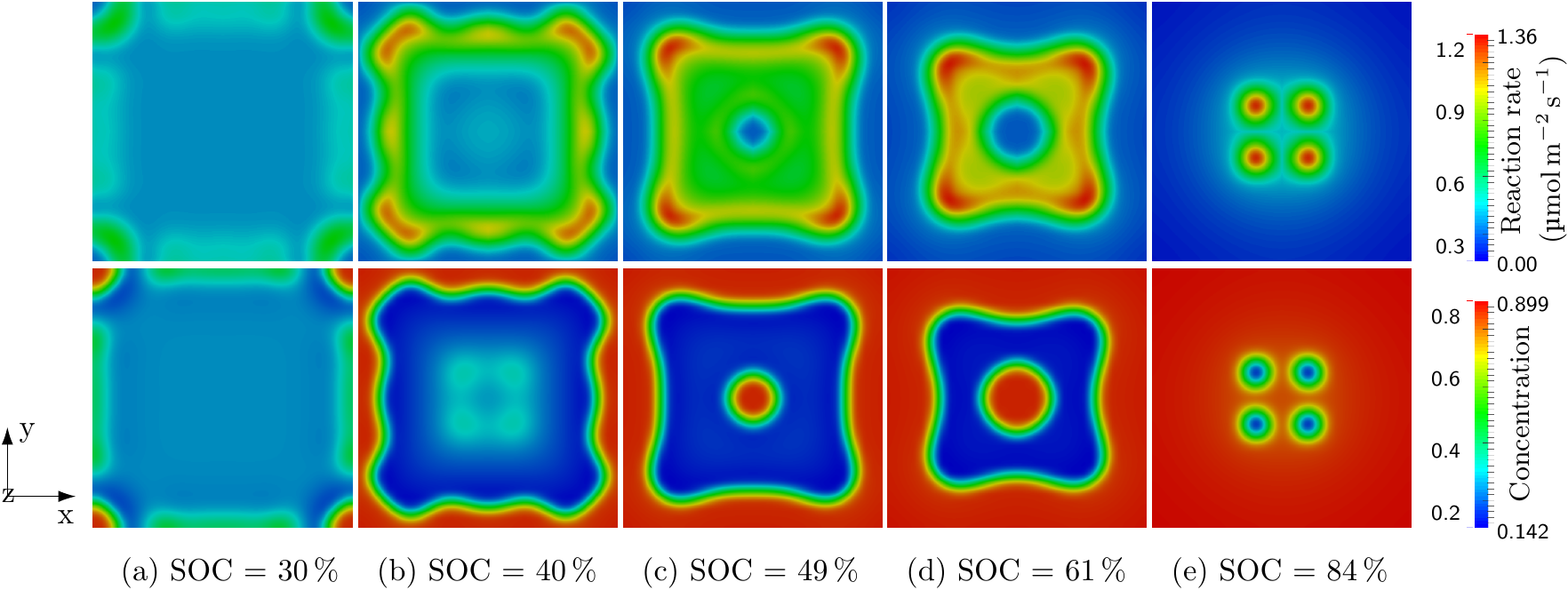}
 \caption{Contour plot of the reaction rate and the concentration SOC in an isotropic diffusion process. The peak
 of the reaction will always take place near the interface.}
 \label{fig:ISO_SP}
\end{figure}

  In the isotropic case, as shown in Figure~\ref{fig:ISO_SP}, the flux is marching towards the interior from all four sides, and 
  the phase segregation of a Li-rich frame and a Li-deficient depression forms.
  As more flux comes in, there arises an island of Li-rich phase in the middle of the Li-poor phase.
  This can be explained by the dynamics of diffusion and reaction, where the whole system is perturbed strongly and it is easy 
  to achieve a phase segregation once the magnitude of fluctuation is large enough.
  As for the reaction, by comparing the concentration profile and the reaction rate, one can observe that the reaction peaks
  at the interface front, where the concentration gradient is very high. 
  In the two homogeneous phases, the reaction is relatively slow, especially in the Li-rich phase, where the reaction almost stops.
  This low efficiency of reaction in the Li-rich phase explains again why the core-shell structure is lithiated much slower than the other, studied in the last subsection. 
  
\begin{figure}[h!]
\centering
\includegraphics[width = \textwidth]{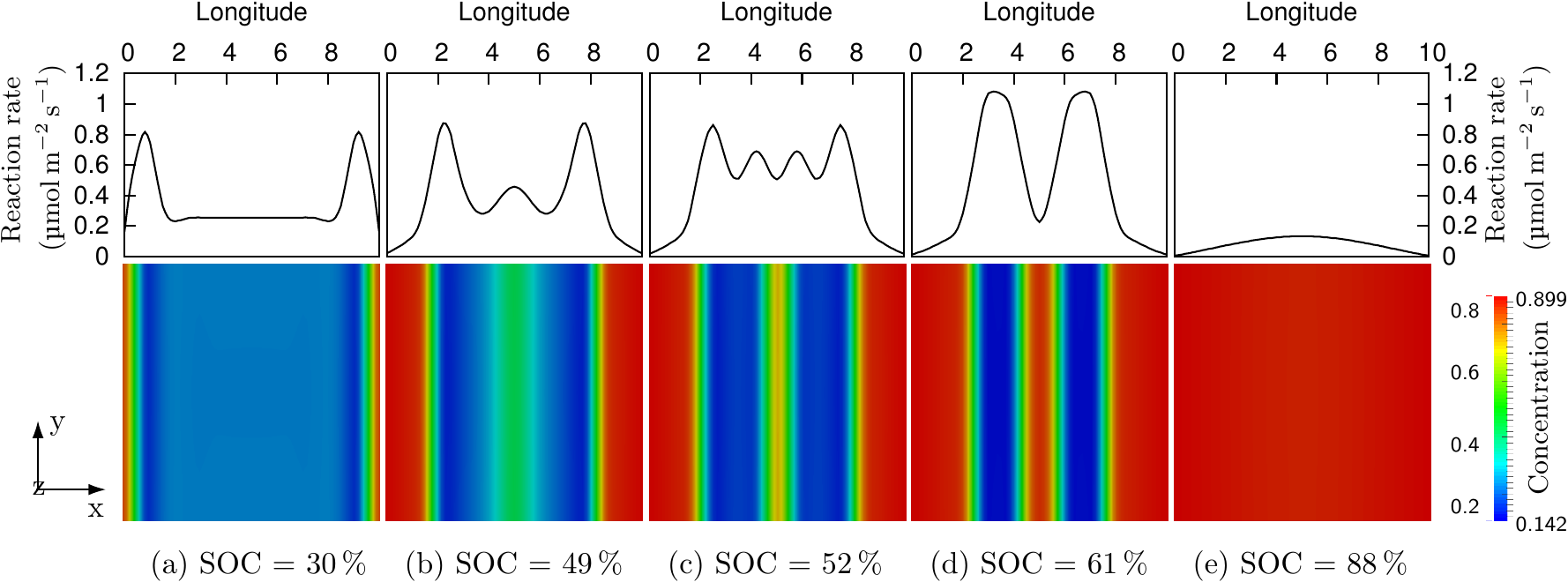} 
\caption{Contour plot of the concentration and plot of the reaction rate along the direction of the 
slowest diffusion (x direction) at different SOC in an anisotropic diffusion process.}
 \label{fig:ANI_SP}
\end{figure}

Figure~\ref{fig:ANI_SP} shows the case of an anisotropic diffusion. As explained in the section \ref{subsec:FM_FreeEnergyDensity},
the interfacial parameter is chosen in such a way that $\kappa_x : \kappa_y : \kappa_z = \mathrm{D}_x : \mathrm{D}_y : \mathrm{D}_z$. In the simulation, the diffusion in x direction is set to be slowest. Note that even though diffusivity in y, z direction is the same, in z-direction lithium sites are filled faster than in y-direction. This is due to the fact that  the dimension in z-direction is smaller than that in y-direction. 
In contrast to the isotropic case, phase segregation initiates from the two ends although the reaction takes place in all six sides
of x axis. As time goes on, the phase interface marches towards the center. A third Li-rich phase appears in the middle when SOC is
approximately $\SI{50}{\percent}$, thus the formation of stripes appears.  This result supports the 
domino-cascade model of LiFePO$_4$, which was proposed in the work of Delma et al.~\cite{Delmas_2008}

These two cases are extreme. However, with proper implementation of crystal anisotropy, 
by filling the diffusion matrix also
in the off-diagonal entries, one can also achieve a core-shell structure with a polygon core,
as observed in the work of Liu et al.~\cite{Liu_2012}.

\subsection{Reaction on the crack surface}\label{subsec:SR_CrackSurface}

\begin{figure}
\begin{floatrow}
\ffigbox{%
  \includegraphics[width =0.38\textwidth]{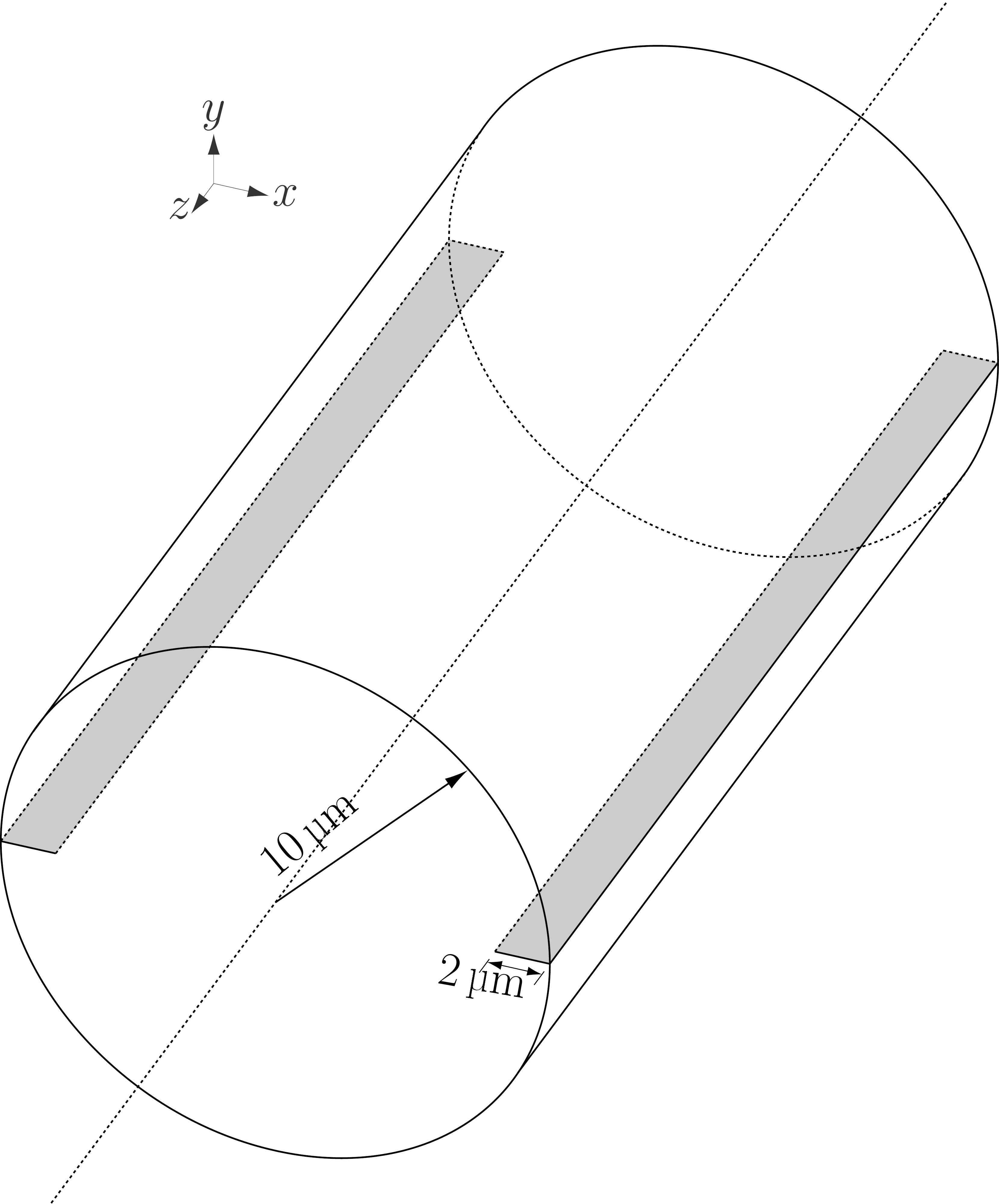}%
}{%
  \caption{Illustration and measurements of an infinite cylinder with initial cracks. One quarter of 
  a disc with a thickness of $\SI{1}{\micro\meter}$ under plane strain assumption is simulated.}%
 \label{fig:Disc_model}
    }
\capbtabbox{%
  \scriptsize
  \begin{tabular}{rl}\hline\hline
     Voltage drop eld./ely. ($\mathbb{\Delta}\phi$)&$\SI{4.88}{\milli\volt}$\\\hline
     Single reac. step time ($\tau_0$)&$\SI{0.01}{\second}$\\\hline
     Initial concentration ($c_0$)&$\SI{0.8}{}$\\\hline
     Energy release rate ($\mathcal{G}_c$)&$\SI{6d-2}{\newton\per\meter}$\\\hline
     Crack lenth scale ($\epsilon$) &$\SI{0.05}{\micro\meter}$\\\hline
     Crack mobility (M$_c$)&$\SI{1.3d-3}{\joule\per\meter\cubed\per\second}$\\\hline\hline
      \end{tabular}
}{%
  \caption{Simulation parameters for the crack propagation problem. Others parameters can be found in
      Table~\ref{tb:parameters}. }%
\label{tb:parameter_Disc}
      }
\end{floatrow}
\end{figure}

As final example, we simulate an infinitely long cylinder with two initial parallel longitudinal cracks on its exterior.
The problem is illustrated in Figure~\ref{fig:Disc_model}, and the corresponding parameters are given
in Table~\ref{tb:parameter_Disc}. The electrode/electrolyte voltage drop is given 
such that Li is extracted from the cylinder.
The reaction only takes place on the cylindrical surface and the crack
surface. As it is explained in section~\ref{subsec:ECR_CrackSurface}, the reaction on the crack surface
is approximated by the weighted source in the phase-field theory. In order to reach the diffusive profile for the initial crack, the reaction is set to be zero for the first 3 seconds. 

\begin{figure}[h!]
 \centering
 \includegraphics[width =\textwidth]{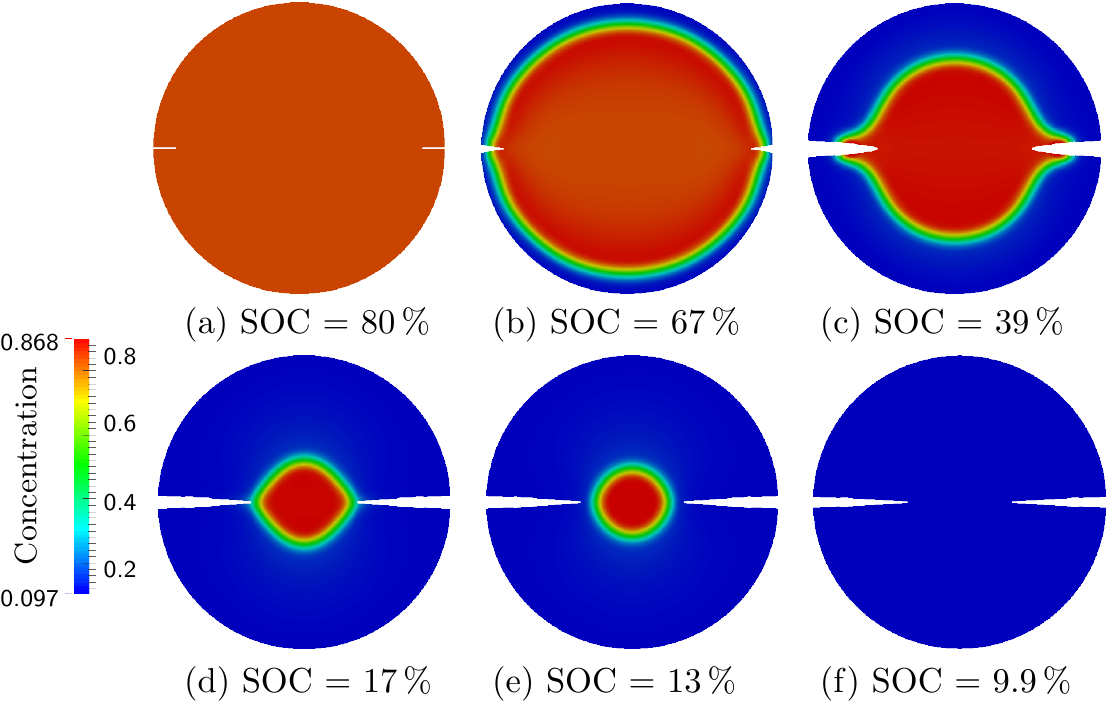}
 \caption{Crack propagation under delithiation and phase segregation. 
 a) Initial homogeneous state. b) Formation of phase segregation initiates
 the crack propagation. c) Intermediate stage when the crack propagates faster than phase interface. d) Stage when phase interface leaves the 
 crack tip. e) Stage when phase interface moves towards center. d) Final stage when the reaction stops.}
 \label{fig:Disc_crack}
\end{figure}

The results of the crack propagation is shown in Figure~\ref{fig:Disc_crack}. Initially, the 
concentration field is homogeneous. As the outer layer loses more lithium, a two-phase profile
appears. It should be noted that at the crack tip lithium can be supplied quickly from
the unbroken material. In fact, due to the large tensile stresses at the crack tip, the drift effect of the  mechanical field towards the crack tip becomes prominent. 
This effect can be seen more clearly in Figure~\ref{fig:Disc_crack}(c), where the 
phase interface on the crack surface is far behind that in the other part of the material.

On the other hand, due to the loss of lithium, the outer layer turns to shrink. Because of this mismatch with the interior Li-rich phase, tensile circumferential stresses arise in the outer layer, which drives the crack to
propagate. At the first stage, the crack propagates faster than the interface. It then slows down,
until the phase interface runs over the crack tip. After the phase interface leaves the crack, the propagation of crack turns to stop, due to the decrease of the driving force. However, the phase interface continues to move. The Li-rich
phase gradually reforms into a circular. At the end of the simulation, the whole material is almost fully delithiated, and the reaction stops. It should be mentioned that the interplay between the phase segregation and the crack propagation can strongly depend on the choice of the kinetics parameters. A more comprehensive study on this topic will be carried out in the future.

\begin{figure}[h!]
 \centering
 \includegraphics[width =0.7\textwidth]{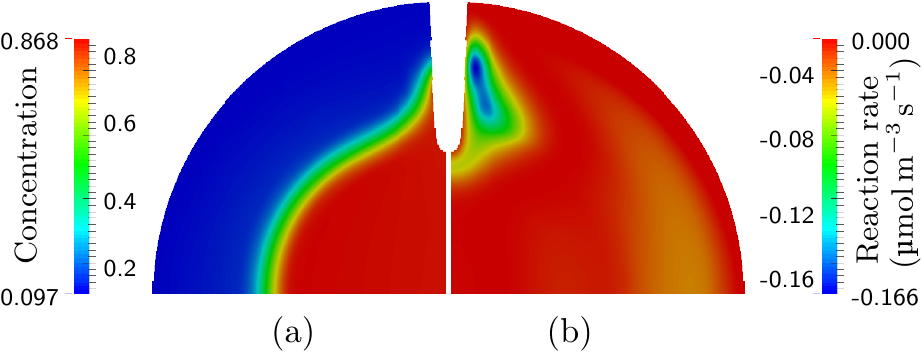}
 \caption{Reaction rate at the crack interface. a) The concentration profile when
 SOC = $\SI{40}{\percent}$. b) The simulated reaction rate around the crack interface.}
 \label{fig:Disc_crack_reac}
\end{figure}

During the whole process, lithium can indeed be released on the crack surface. 
To check the approximated reaction on the crack surface, the reaction and the corresponding
concentration profile on the circular plate at SOC = $\SI{40}{\percent}$ are demonstrated in Fig.~\ref{fig:Disc_crack_reac}. At this state, on the crack
surface both phases and phase interface are exposed to the electrolyte.
By comparing these two plots, it can be seen that near the crack surface, 
the reaction peaks in the neighborhood of the interface  
and gradually diffuses into the bulk. In the Li-poor region, the reaction almost 
vanishes, while towards the crack tip the reaction is rather strong.

\section{Conclusion}
In this paper, the electrochemical reaction in the lithium ion batteries
is studied by using phase-field modeling of fracture coupled with anisotropic Cahn-Hilliard-type diffusion
in the large deformation regime.

The reaction on the surface is modeled through a modified Butler-Volmer equation, taking into account
the influence of phase separation on the surface. The reaction on the crack surface is considered
as a source term within the volume weighted by a damage-variable-related term to constrain the reaction to take place
only on the 
transition zone between the unbroken and broken state.

Three examples are carried out to study different aspects.
The first example of an isotropic sphere shows that the ratio of the timescale of the reaction and 
diffusion greatly influences the phase segregation of the material: a fast reaction gives 
a ``shrinking core'' while a slow reaction has the nucleation initiated already from the surface.
In turn, the phase segregation also influences the real reaction rate through the electrochemical potential
on the surface. When a core-shell structure is formed, the highly homogeneous concentration on the surface
prevents the lithium from further inserting into the particle. However, an uneven distribution of the
concentration, although accompanied by a highly distorted surface, can give a much more robust reaction in
the long run.
In the second example, the reaction on the interface of two phases in an isotropic diffusion and anisotropic 
diffusion is studied. Results show that, both in isotropic and anisotropic cases, the reaction rate peaks near
the interface, where there exists a large concentration gradient.
The last example shows the reaction on the crack surface. It is shown that the crack evolution can be driven
by a outflow of the species when the material exhibits a phase segregation behavior. The electrochemical reaction on the newly created crack surfaces has been discussed, along with the interaction between the crack propagation and the phase segregation process.

\section*{Acknowledgments}

This work is supported by the ``Excellence Initiative'' of the German Federal
and State Governments and the
Graduate School of Computational Engineering at the Technische Universit\"at Darmstadt.


\end{document}